\documentclass[12pt, draftclsnofoot, onecolumn]{IEEEtran}

\IEEEoverridecommandlockouts

\usepackage{graphicx,cite,acronym,amsmath,amssymb,amsfonts,color,subfigure,floatflt,stfloats}
\usepackage{epsfig,epstopdf,psfrag}
\usepackage[ruled,vlined]{algorithm2e}
\SetKwInput{KwInput}{Input}                
\SetKwInput{KwOutput}{Output}  

\usepackage[table]{xcolor}
\usepackage{bm}
\usepackage[level=3]{wgroup_message}  

\usepackage[cmintegrals]{newtxmath}
\usepackage{graphics,hhline,array,cite,bbm}

\usepackage{latexsym}
\usepackage{theorem}
\usepackage{times}
\usepackage{makeidx}

\DeclareGraphicsExtensions{.png,.jpg,.eps}

\acrodef{CR}{channel response}

\acrodef{BS}{base station}

\acrodef{MS}{mobile station}

\acrodef{UE}{user equipment}

\acrodef{MIMO}{multiple-input multiple-output}

\acrodef{RIS}{reconfigurable intelligent surface}

\acrodef{LIS}{large intelligent surface}

\acrodef{MIS}{medium intelligent surface}

\acrodef{SIS}{small intelligent surface}

\acrodef{DoF}{degrees-of-fredom}

\acrodef{AF}{amplify \& forward}

\acrodef{DF}{detect \& forward}

\acrodef{JF}{just forward}

\acrodef{CSI}{channel state information}

\acrodef{RV}{random variable}

\acrodef{i.i.d.}{independent, identically distributed}

\acrodef{PSD}{power spectral density}

\acrodef{PDF}{probability distribution function}

\acrodef{CDF}{cumulative distribution function}

\acrodef{ch.f.}{characteristic function}

\acrodef{AWGN}{additive white Gaussian noise}

\acrodef{RSSI}{received signal strength indicator}

\acrodef{SNR}{signal-to-noise ratio}

\acrodef{LRT}{likelihood ratio test}

\acrodef{GLRT}{generalized likelihood ratio test}

\acrodef{GML}{generalized maximum likelihood}

\acrodef{LOS}{line-of-sight}

\acrodef{NLOS}{non-line-of-sight}

\acrodef{GDOP}{geometric dilution of precision}

\acrodef{GPS}{Global Positioning System}

\acrodef{FIM}{Fisher information matrix}

\acrodef{PEB}{position error bound}

\acrodef{WSN}{Wireless Sensor Network}

\acrodef{MAC}{medium access control}

\acrodef{RSS}{received signal strength}

\acrodef{RTT}{round-trip time}

\acrodef{MIMO}{multiple-input multiple-output}

\acrodef{MF}{matched filter}

\acrodef{ED}{energy detector}

\acrodef{ML}{maximum likelihood}

\acrodef{NL}{nonlinear}

\acrodef{MSE}{mean square error}

\acrodef{RMSE}{root mean square error}

\acrodef{ppm}{part-per-million}

\acrodef{PRP}{pulse repetition period}

\acrodef{ACK}{acknowledge}

\acrodef{UWB}{ultrawide bandwidth}

\acrodef{TNR}{threshold-to-noise ratio}

\acrodef{NLOS}{non line-of-sight}

\acrodef{LOS}{line-of-sight}

\acrodef{LS}{least squares}

\acrodef{IR-UWB}{impulse radio UWB}

\acrodef{FCC}{Federal Communications Commission}

\acrodef{TH}{time-hopping}

\acrodef{PPM}{pulse position modulation}

\acrodef{PAM}{pulse amplitude modulation}

\acrodef{MUI}{multi-user interference}

\acrodef{PDP}{power delay profile}

\acrodef{PPP}{Poisson point process}

\acrodef{DS}{delay spread}

\acrodef{CED}{channel excess delay}

\acrodef{BPZF}{band-pass zonal filter}

\acrodef{SIR}{signal-to-interference ratio}

\acrodef{RFID}{radio frequency identification}

\acrodef{WPAN}{wireless personal area networks}

\acrodef{WWLB}{Weiss-Weinstein lower bound}

\acrodef{DP}{direct path}

\acrodef{MF}{matched filter}

\acrodef{MMSE}{minimum-mean-square-error}

\acrodef{SBS}{serial backward search}

\acrodef{NBI}{narrowband interference}

\acrodef{WBI}{wideband interference}

\acrodef{INR}{interference-to-noise ratio}

\acrodef{CIR}{channel impulse response}

\acrodef{ISI}{inter-symbol interference}

\acrodef{CPR}{channel pulse response}

\acrodef{LRT}{likelihood ratio test}

\acrodef{RADAR}{RADAR}

\acrodef{MUR}{Multistatic RADAR}

\acrodef{MUI}{multi-user interference}

\acrodef{e.m.}{electromagnetic}

\acrodef{CW}{continuous wave}

\acrodef{RF}{radiofrequency}

\acrodef{FCC}{Federal Communications Commission}

\acrodef{EIRP}{effective radiated isotropic power}

\acrodef{RCS}{radar cross section}

\acrodef{BAV}{balanced antipodal Vivaldi}

\acrodef{PRake}{partial Rake}

\acrodef{RTLS}{real time locating system}

\acrodef{CRB}{Cram\'{e}r-Rao bound}

\acrodef{ZZB}{Ziv-Zakai bound}

\acrodef{TOA}{time-of-arrival}

\acrodef{TOF}{time-of-flight}

\acrodef{WSN}{wireless sensor network}

\acrodef{MAC}{medium access control}

\acrodef{RSS}{received signal strength}

\acrodef{TDOA}{time difference-of-arrival}

\acrodef{RF}{radiofrequency}

\acrodef{RTT}{round-trip time}

\acrodef{AOA}{angle-of-arrival}

\acrodef{MF}{matched filter}

\acrodef{ED}{energy detector}

\acrodef{ML}{maximum likelihood}

\acrodef{MUR}{Multistatic radar}

\acrodef{HDSA}{high-definition situation-aware}

\acrodef{RRC}{root raised cosine}

\acrodef{OFDM}{orthogonal frequency division multiplexing}

\acrodef{IF}{intermediate frequency}

\acrodef{PHY}{physical layer}

\acrodef{S-V}{Saleh-Valenzuela}

\acrodef{UHF}{ultra-high frequency}

\acrodef{PR}{pseudo-random}

\acrodef{SoC}{System on Chip}

\acrodef{SoP}{System on Package}

\acrodef{SPMF}{Single-Path Matched Filter}

\acrodef{IMF}{Ideal Matched Filter}

\acrodef{SCR}{signal-to-clutter ratio}

\acrodef{BEP}{bit error probability}

\acrodef{BER}{bit error rate}

\acrodef{WSR}{wireless sensor radar}

\acrodef{HPBW}{half power beam width}

\acrodef{LEO}{localization error outage}

\acrodef{WSS}{wide-sense stationary}

\acrodef{TE}{transverse eletric}

\acrodef{TR}{time-reversal}

\acrodef{WSSUS}{WSS with uncorrelated scattering}

\acrodef{GP}{Gaussian process}

\acrodef{IMU}{inertial measurement unit}

\newcommand{\EX}[1] {{\mathbb{E}}\left\{{#1}\right\}}

\newcommand{\boldp} {{\bf{p}}}

\newcommand{\0}{\mathbf{0}}

\newcommand{\SNR}{\text{SNR}}

\newcommand{\boldpbs} {{\bf{p}_{\text{BS}}}}

\newcommand{\fr} {f_{\text{r}}}

\newcommand{\Lx} {L_{\text{x}}}
\newcommand{\Ly} {L_{\text{y}}}

\newcommand{\Ptx} {P_{\text{T}}}

\newcommand{\Gs} {G_{\text{S}}}
\newcommand{\Gt} {G_{\text{T}}}
\newcommand{\Gr} {G_{\text{R}}}

\newcommand{\thetai} {\theta_{i}}
\newcommand{\phii} {\phi_{i}}

\newcommand{\dfraun} {d_{\text{Fraunhofer}}}
\newcommand{\dfresn} {d_{\text{Fresnel}}}

\newcommand{\Gc} {{  G_{\text{c}}}}

\newcommand{\Ac} {{  A_{\text{c}}}}

\newcommand{\Thetai} { \Theta_{\text{i}}}

\newcommand{\ux} { u_{x}}
\newcommand{\uy} { u_{y}}

\begin{document}
\title{Using MetaPrisms for Performance Improvement in Wireless Communications}

\author{
\IEEEauthorblockN{Davide~Dardari,~\IEEEmembership{Senior~Member,~IEEE}, Devis Massari}
\IEEEcompsocitemizethanks{\IEEEcompsocthanksitem 
 D.~Dardari is  with the 
   Dipartimento di Ingegneria dell'Energia Elettrica e dell'Informazione ``Guglielmo Marconi"  (DEI), CNIT, 
   University of Bologna, Cesena Campus, via dell'Universit\'a 52,
   Cesena (FC), Italy, (e-mail: davide.dardari@unibo.it).
    }
    \IEEEcompsocitemizethanks{\IEEEcompsocthanksitem 
    D.~Massari is with Unitec S.p.A, Via Provinciale Cotignola, 20/9 - 48022 Lugo (RA), Italy, (devis.massari@gmail.com).
    }
}


%


\maketitle
\begin{abstract}

In this paper, we put forth the idea of \emph{metaprism}, a passive and non-reconfigurable metasurface acting as a metamirror with frequency-dependent 
reflecting properties within the signal bandwidth. 
We show that, with an appropriate design of the metaprism, it is possible to control that each data stream in an \acf{OFDM} system is reflected in the desired direction without the need for control channels and \acf{CSI} estimation between the base station and the metaprism,  but simply by correctly assigning subcarriers to users.
Furthermore, the metaprism can also be designed so that it focuses the signal towards a specific position depending on the subcarrier,  provided that it is in the near-field, with consequent path-loss reduction. 
A critical discussion is also presented about the path-loss gain obtainable from metaprisms and, more generally, from metasurfaces.
The numerical results show that this solution is surprisingly effective in extending  the coverage in areas experiencing severe \acf{NLOS} channel conditions, thus making it a very appealing alternative to reconfigurable metasurfaces when low-cost, no energy consumption, and backward compatibility with existing wireless standards are required.

\end{abstract}



\section{Introduction}

The exponential growing in traffic demand in wireless networks has forced the exploitation of new frequency bands in the millimeter wave region and, more in perspective, at THz frequencies. At the same time, new performance indicators, such as spatial capacity density, reliability and latency, are becoming increasingly critical in next generation networks \cite{ShaMol:17,Zha:19,RapXinKanJuMadManAlkTri:19}.

When switching to high frequencies, more bandwidth is available on the one hand, but on the other hand, the wireless channel suffers a higher path loss that can be countered by using directional antennas such as antenna arrays. The main disadvantage in the use of high directivity antennas is the greater susceptibility to signal blocking in the presence of obstacles.
In fact, wireless propagation relies mainly  on the presence of the \ac{LOS} direct path  with scarce multipath components
\cite{
 RapXinMacMolMelZha:17}. This makes the coverage of \ac{NLOS} areas more challenging than at the lower frequency bands, where \ac{NLOS} communication can be guaranteed by exploiting the rich multipath deriving from \ac{e.m.} scattering, especially when using massive \ac{MIMO} systems capable of ``focusing" multipath components on receiver position \cite{San:19}.

One possible solution to increase the coverage at millimeter waves is to  add more \acp{BS} or introduce regenerative or non-regenerative relaying nodes \cite{HuaYiZhu:04,DecGueConDErSibDar:14}. However, in many applications or scenarios, relays could be expensive in terms of cost, complexity, and deployment constraints. Moreover, relays introduce additional communication delay (thus in contrast with the need to lower the latency), are not energy efficient, require some coordination with the \ac{BS} (e.g., synchronization), and support only half-duplex communications. Despite the recent introduction of full-duplex relays, the other drawbacks remain \cite{MinPen:14}. 

Another solution is to use RF passive mirrors/reflectors, which  have been used since the introduction of wireless communications, for instance, as gap fillers in broadcast systems to cover areas shadowed by mountains. Their main limitation is that they are not flexible because not programmable. 

The introduction of  metamaterials  to realize, for instance, the so called \emph{metasurfaces}  has boosted a fertile research area \cite{HolKueGorOHaBooSmi:12,Gly:16,Bur:16,GonMinChaMac:2017}.
In fact, with metasurfaces \ac{e.m.} waves can be shaped almost arbitrarily to obtain a given functionality.  
There exists a rich literature describing their possible realizations, considering patches of simple geometry and design for the realization of holographic metasurfaces 
 or many other solutions \cite{Tre:15,Bur:16,HolKueGorOHaBooSmi:12,Hun:14,HolMohKueDie:05,Gly:16,AsaAlbTcvRubRadTre:16}. 
Notably, several of metasurfaces functionalities can be realized with layers of bulk metamaterials, and the possible applications include, but are not limited to, transmitarrays  \cite{DiP:17},
metamirrors \cite{RadAsaTre:14,
AsaAlbTcvRubRadTre:16,AsaRadVehTre:15}, reflectarrays 
\cite{HumCar:14,NayYanEsh:15} and holograms \cite{Gly:16,Hun:14}. 
Along a different direction, a few papers  analyze the potential of using metasurfaces as \ac{LIS} antennas to improve the communication capacity \cite{HuRusEdf:18,NepBuf:17,Eldar:19,Dar:19,BjoSan:20} or to enable single-anchor localization \cite{GuiDar:19}. 

The recent availability  of programmable metasurfaces  to design smart \ac{e.m.} reflectors  using thin meta-materials has opened new very appealing perspectives \cite{Tur:14,Aky:18}. 
These \acp{RIS} can be embedded in daily life objects such as walls, clothes, buildings, etc., and can be used as distributed platforms to perform low-energy and low-complexity sensing, storage and analog computing.   
Environments coated with intelligent surfaces constitute the recently proposed \emph{smart radio environments} concept \cite{LiaNieTsiPitIoaAky:18,
DiRenzo:19b}. In smart radio environments, the design paradigm is changed from wireless devices/networks that adapt themselves to the environment (e.g., propagation conditions), to the joint optimization of both devices and environment using \acp{RIS}. 

The advantages of \ac{RIS}-enabled environments have been analyzed in several papers. For instance, in \cite{YanZhaZha:19} a
\ac{RIS}-enhanced \ac{OFDM} system is investigated, where the power allocation and the  metasurface phase profile are jointly optimized to boost the achievable rate of a cell-edge user. 
In \cite{OzdBjoLar:20}, it is shown how the rank of  \ac{MIMO} communication in \ac{LOS}  can be increased by adding a \ac{RIS} generating an artificial path. 
The authors in \cite{DiRenzo:19c} present a comparison between \ac{RIS}- and relay-enabled wireless networks by discussing the similarities and differences.
Other studies can be found, for instance, in \cite{YeGuoAlo:19,ZhaZha:19,JunSaaDebHon:19}.
     
%
%
%
%
%
%
%

Despite being a promising solution, \ac{RIS}-based systems have some disadvantages which could make them less appealing in several applications. In fact, to reconfigure  a \ac{RIS}  in real-time, a dedicated control channel is needed which might entail a certain signaling overhead and, above all,  additional complexity and cost. Moreover a \ac{RIS} needs to be powered and this might not be possible or convenient in many scenarios. 
A fundamental challenge when using a \ac{RIS} is the estimation of the \ac{CSI} \cite{DiRenzo:19b}. In fact, the \ac{CSI} related to the \ac{BS}-\ac{RIS}  and the \ac{RIS}-user links must be estimated and available to the network coordinator in order to configure properly the phase profile of the \ac{RIS} through the control channel. While the \ac{BS}-\ac{RIS} link can be estimated considering that in many applications the position of the \ac{RIS} with respect to the \ac{BS} is known, the \ac{RIS}-user link is more challenging to estimate as the location of users changes as well as the environment conditions.  Again, this means higher \ac{RIS} complexity, cost, signaling overhead, and a dedicated wireless technology which might not be compliant with existing standards, making \acp{RIS} not always competitive with relays \cite{DiRenzo:19c}. 

\begin{figure}[t]
\centerline{\includegraphics[width=0.8\columnwidth,angle=0]{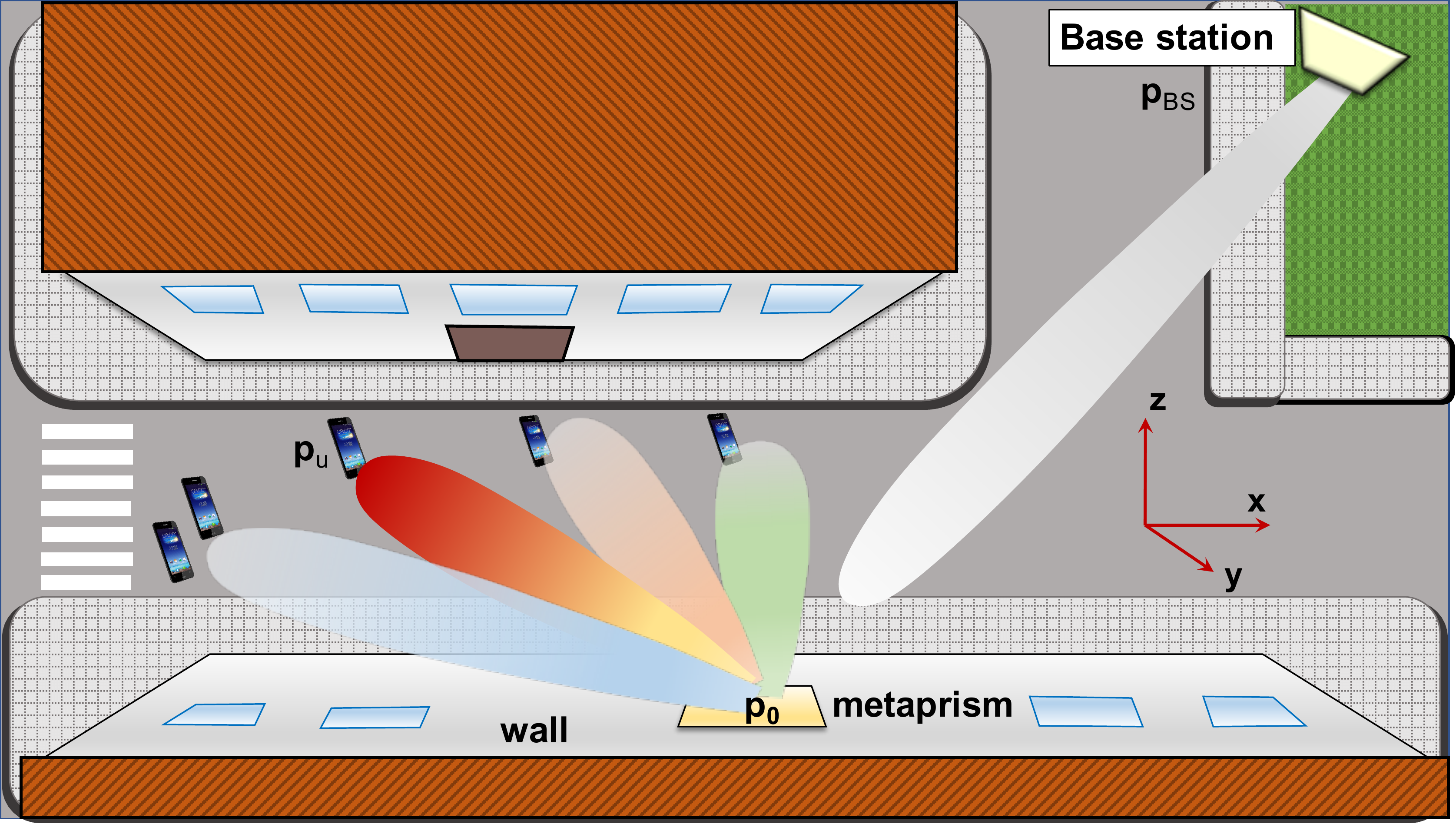}}
\caption{The considered \ac{NLOS} scenario empowered by a metaprism. }
\label{Fig:Scenario}
\end{figure} 

In this paper, we introduce the novel idea of \emph{metaprism}, a passive and non-reconfigurable metasurface that acts as a metamirror, whose reflecting properties are frequency-dependent within the signal bandwidth. 
A metaprism can be introduced in a scenario where users are in \ac{NLOS} condition with respect to the \ac{BS} in order to extend  the covered area at low cost (see the scenario example in Fig. \ref{Fig:Scenario}). 
Thanks to the metaprism, one can control the reflection of the signal through a proper selection of the subcarrier  assigned to each user
using a conventional  \ac{OFDM} signaling, without interacting with the metaprism and without the need for \acf{CSI} estimation between the users and the metaprism.
We show that, with an appropriate design of the frequency-dependent phase profile of the metaprism,  a significant path-loss reduction can be obtained by steering/focusing the signal towards/on a specific position depending on the assigned subcarrier. 

The design of frequency-selective surfaces (FSS) is not new: for instance, they are used in  dual-band reflectarrays \cite{TayTanPalUdpUdpRot:15}, antenna covering surfaces, frequency-selective absorbers and, in general, to perform spectral filtering in both microwave and optical ranges through the design of metasurfaces with extremely dispersive reflection or transmission properties \cite{Gly:16}. 
However, to the knowledge of the Authors, this is the first paper proposing and analyzing how to exploit  frequency-selective properties of metasurfaces to improve the coverage of short-range wireless networks.  

The numerical results corroborate the validity of the idea showing the significant performance improvement in wireless network coverage and achievable rate, making it a very appealing alternative to \acp{RIS}.
In particular, the main advantages of a metaprism with respect to a \ac{RIS}  or a relay are:
\begin{itemize}
\item It is completely passive and hence it does not require any power supply;
\item There is no need neither for control channel nor for \ac{CSI} estimation of each link;
\item It does not introduce any reconfiguration delay as reflection properties depend on the impinging signal (subcarrier);
\item It is transparent to the communication technology, i.e., it can be applied also to current wireless standards based on \ac{OFDM} to enhance the coverage;
\item Being less complex and completely passive, it is expected to facilitate its widespread diffusion and reduce costs.  
\end{itemize} 

Like \acp{RIS}, metaprisms are intrinsically full-duplex and introduce zero-latency. 

The remainder of this paper is organized as follows. In Section~\ref{Sec:Metasurface}, the frequency-dependent modeling of metasurface behavior is introduced.  
The \ac{OFDM} wireless link aided by metaprisms is characterized in Section~\ref{Sec:SystemModel}, whereas in Sections~\ref{Sec:Beamsteering}
and \ref{Sec:DesignFocusing}, the design criteria of phase profiles for the metaprism to realize, respectively,  frequency-dependent beamsteering and focusing are elaborated. 
Some considerations about the validity of path-loss models  when using metaprisms or, more in general, \acp{RIS} are given in Section~\ref{Sec:PL}. 
An example of algorithm to assign subcarriers in a multi-user scenario in order to equalize and maximize the per-user achievable rate is proposed in  Section~\ref{Sec:AR}.
Numerical results and discussions are presented in Section~\ref{Sec:NumericalResults}. 
Finally, conclusions are drawn in Section~\ref{Sec:Conclusion}.

%
%
%

\section{Frequency-selective Models for Metasurfaces}
\label{Sec:Metasurface}

\subsection{General Model}

\begin{figure}[t]
\centering
\subfigure[][ Metasurface composed of elementary cells. Equivalent model of the cell.]{\label{Fig:EquivalentModel1}\includegraphics[width=0.51\textwidth]{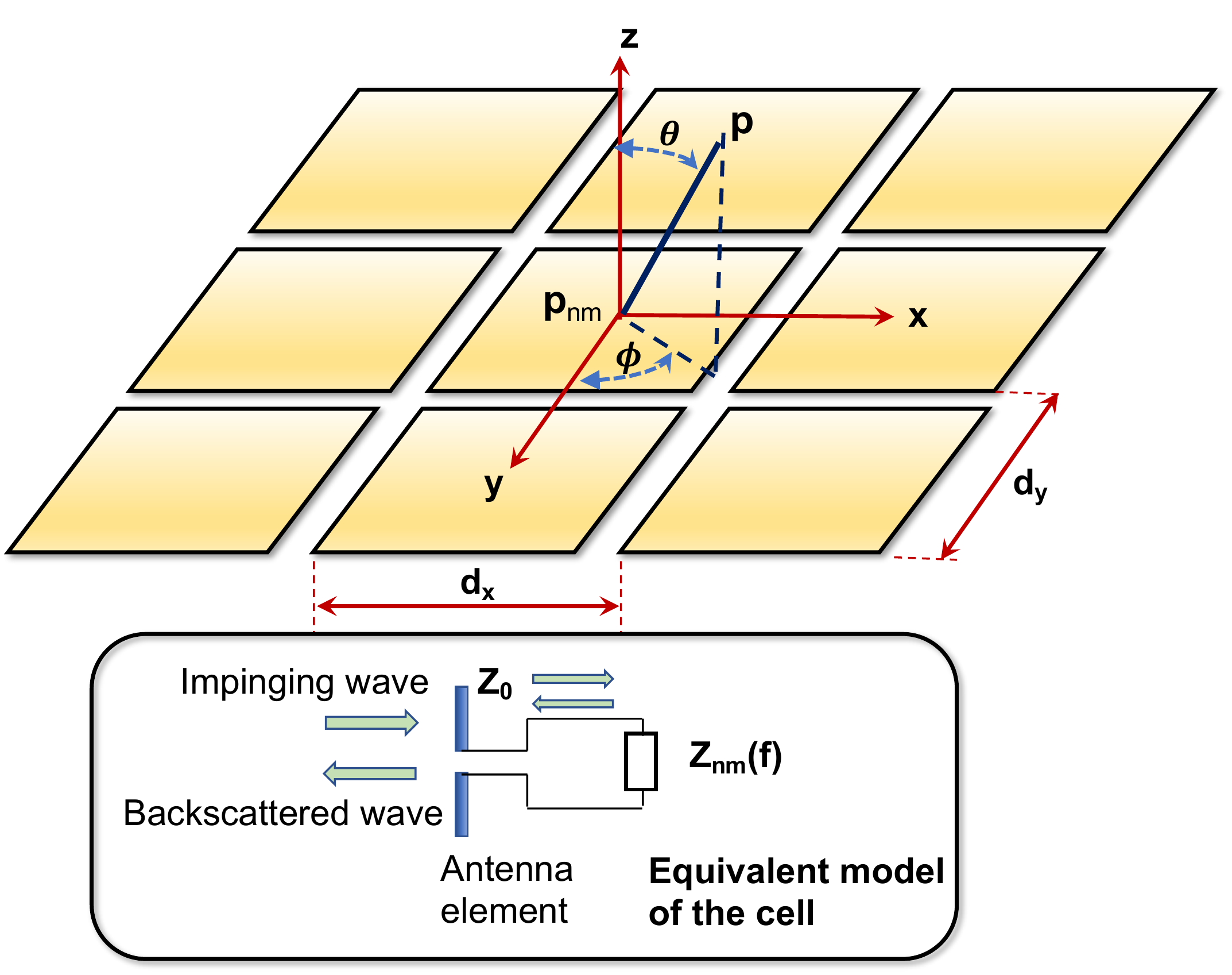}}
\subfigure[][Thin metasurface composed of omega-shaped particles \cite{RadAsaTre:14}. ]{\label{Fig:EquivalentModel2}\includegraphics[width=0.48\textwidth]{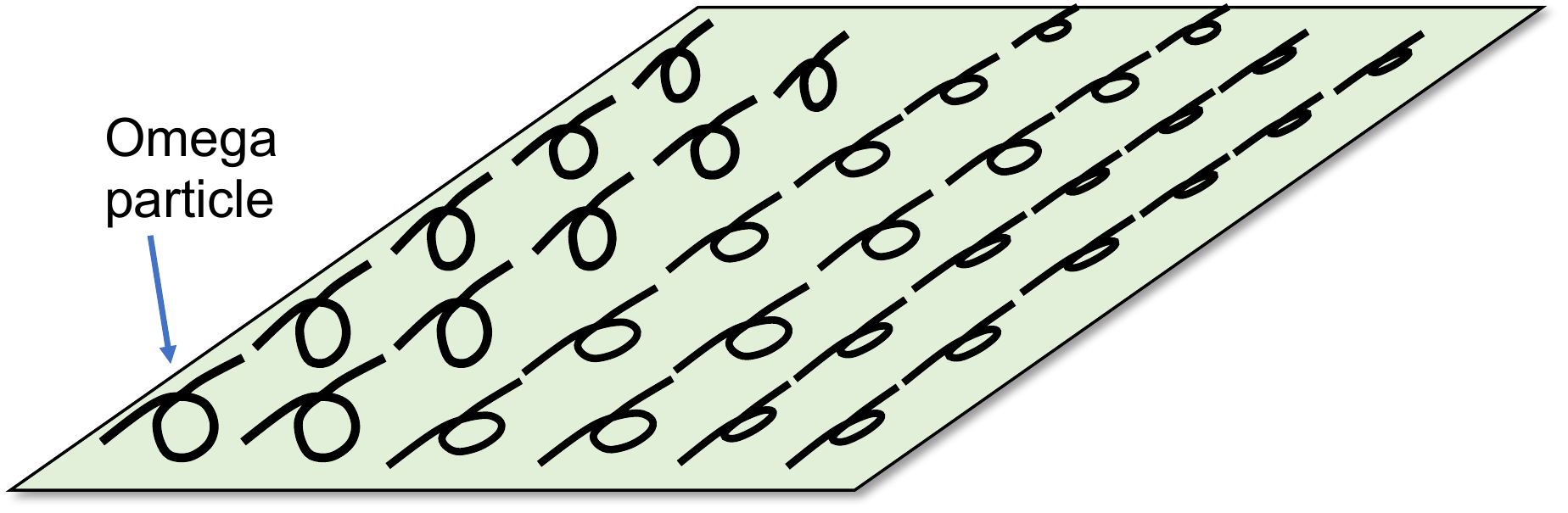}}
\caption{Examples of metasurfaces.}
\vskip -0.3cm
\end{figure}

With reference to Fig. \ref{Fig:EquivalentModel1}, consider a metasurface in the $x-y$ plane with center at coordinates $\boldp_0=(0,0,0)$, consisting of $N \times M$ cells distributed in a grid  of points with coordinates $\boldp_{nm}=(x_n,y_m,0)$, where  $x_n=n\,  d_x-L_x/2 $, $n=0,1,\ldots N-1$, and $y_m=m\,  d_y-L_y/2 $, $m=0,1,\ldots M-1$, being $L_x=N\, d_x$ and $L_y=M\, d_y$ the surface's size. The cell spacing   $d_x$, $d_y$, respectively, in the $x$ and $y$ directions, are often much smaller than the wavelength $\lambda$, typically $d_x, d_y \approx \lambda/2 -\lambda/10$ \cite{Bur:16,HolKueGorOHaBooSmi:12}.

There are several technologies to realize a metasurface each of them obeys to a specific model. A rough classification can be done between metasurfaces whose cells can be seen as small radiating elements with tunable load impedance \cite{TanCheDaiHanDiRZenJinCheCui:19,Ell:19,HumCar:14,SmiOkaPul:17}, i.e., using volumetric metamaterials with several wavelength thick, and subwavelength metasurfaces producing a modification of the \ac{e.m.} field which can be modeled as an impedance sheet \cite{PfeGrb:13,SelEle:13,Gly:16,MohNasAnd:16,AsaAlbTcvRubRadTre:16}.

For the first class, a quite general equivalent model of the single $nm$th cell of the metasurface is shown in Fig. \ref{Fig:EquivalentModel1}, which consists of a radiation element (antenna) above a ground screen loaded with a cell-dependent  impedance $Z_{nm}(f)$.
The impedance is designed in such a way it is not matched to the antenna impedance $Z_0$, 
thus determining a reflected wave which is irradiated back by the radiation element.
The corresponding  frequency-dependent reflection coefficient in the presence of an incident plane wave with 2D angle $\Thetai=(\thetai,\phii)$ and observed at angle $\Theta=(\theta,\phi)$ is\footnote{We adopt the conventional  spherical coordinate system where $\phi \in [0,2\pi)$ (azimuth) and $\theta \in [0,\pi)$ (inclination).}   \cite{TanCheDaiHanDiRZenJinCheCui:19,Ell:19}
\begin{align} \label{eq:rnm1}
 r_{nm}(\Thetai,\Theta;f)=\sqrt{F(\Thetai) \, F(\Theta)} \, \Gc \, \Gamma_{nm}(f) = \beta_{nm}(\Thetai,\Theta;f) \, e^{\jmath \,  \Psi_{nm}(f)} \, ,
\end{align}
where $F(\Theta)$ is the normalized power radiation pattern that accounts for possible non-isotropic behavior of the radiation element, which we consider,  as first approximation,  frequency-independent within the bandwidth of interest, $\Gc$ is the boresight antenna gain, $\Gamma_{nm}(f)$ is the load reflection coefficient, $\beta_{nm}(\Thetai,\Theta;f)$ is the reflection amplitude and $\Psi_{nm}(f)$ is the reflection phase.\footnote{A more rigorous model should also account for the signal reflected back by the antenna according to its structural \ac{RCS} component \cite{BalB:16}.}
For instance, in \cite{Ell:19} the following parametric shape for $F(\Theta) $ is proposed 
 \begin{align} \label{eq:F}
    F(\Theta)=\left\{\begin{array}{ll} \cos^q (\theta)  & \, \, \, \, \, \theta \in [0,\pi/2] \, , \phi \in [0,2\pi]  \\0 & \, \, \, \, \, \text{otherwise}  \\ \end{array}\right.   \, .   
 \end{align}
 Parameter $q$ depends on the specific technology adopted as well as on the dimension of the cell and it is related to the  boresight gain $\Gc=2\, (q+1)$. Following a similar approach as in \cite{Ell:19}, one possibility is to set $\Gc$ so that the effective area of the cell is equal to the area of the cell $\Ac=d_x \, d_y$, i.e., $\Gc=\Ac\,  4\pi /\lambda^2$, assuming an ideal radiation efficiency. Considering a cell with $d_x=d_y=\lambda/2$, it follows that $\Gc=\pi \simeq 5\,$dBi, and $q=0.57$.  
 A similar model is presented in \cite{TanCheDaiHanDiRZenJinCheCui:19} with $q=3$. 
 %
 %
The load reflection coefficient is given by 
\begin{equation} \label{eq:Gamma}
\Gamma_{nm}(f) = \frac{Z_{nm}(f)-Z_0}{Z_{nm}(f)+Z_0} \, .
\end{equation} 
%
By properly designing  the impedance $Z_{nm}(f)$ at each cell it is possible to realize different reflecting behaviors of the metasurface as it will be illustrated in the next section.     

Examples related to the second class of metasurfaces consisting in a very thin surface 
can be found, for instance, in \cite{PfeGrb:13,SelEle:13,RadAsaTre:14,Gly:16,AsaRadVehTre:15,HolMohKueDie:05,MohNasAnd:16,AsaAlbTcvRubRadTre:16}. 
 Specifically, if one imposes that the metasurface acts as a reflector  with no transmission (metamirror), the boundary conditions  on the surface, necessary to satisfy the Maxwell's equations, require the presence of  electric and magnetic currents. Electrically and magnetically polarizable metasurfaces with thickness much less than the wavelength $\lambda$ can be composed by small 
 orthogonal electric and magnetic dipoles tangential to the surface, 
 thus realizing a dense set of Huygens sources (meta-atoms). For example, each meta-atom can be realized using a omega-shaped particle as shown in Fig. \ref{Fig:EquivalentModel2} \cite{RadAsaTre:14}.   

According to the specific technology employed, the position-dependent  impedance of the surface (impedance sheet)  can be designed   
to obtain the desired reflection coefficient in the form as in the right hand side of  \eqref{eq:rnm1}.
%
Examples on how impedance  characteristic is related to the surface's impedance can be found in \cite{RadAsaTre:14}.
It has been demonstrated that using local passive loss-less surfaces there is always a power loss during the reflection. Instead, 
no power loss is possible by allowing the surface to be  locally non-passive even though globally passive, i.e., allowing periodical flow of power into the metasurface structure and back \cite{MohNasAnd:16,RadAsaTre:14,AsaAlbTcvRubRadTre:16}.   

Regardless the specific technology adopted, as it will be clearer later, in general we aim to design the metasurface such that the reflection phase shift could be put, at least as first approximation, in the following form 
\begin{align} \label{eq:Psif}
 \Psi_{nm}(f) \simeq \alpha_{nm} \cdot (f-\fr) + \gamma(f) \, , 
\end{align}
for $f$ within the signal bandwidth, where $\alpha_{nm}$ is a  cell-dependent coefficient and $\gamma(f)$ is a (possibly present) frequency-dependent phase shift.
 In particular, $\gamma(f)$ represents a common (among cells) phase offset which is irrelevant to beamsteering and focusing operations.  For this reason in the remaining text we will neglect it.
According to the desired reflection behavior, the reference frequency $\fr$ can be chosen either equal to the signal center frequency $f_0$ or equal to the lowest  frequency edge of the signal band. 

As it will be evident in the next section, the reflection characteristics of a metasurface designed with the phase profile \eqref{eq:Psif} depend on the frequency; for this reason we name it \emph{metaprism}. 


\subsection{Design Example}

We illustrate an example of how the phase response \eqref{eq:Psif} can be obtained starting from the equivalent model in Fig. \ref{Fig:EquivalentModel1} described by \eqref{eq:rnm1}, \eqref{eq:F}, and \eqref{eq:Gamma}.  
We consider a purely reactive impedance $Z_{nm}(f)=\jmath X_{nm}(f)$ and a purely resistive antenna impedance $Z_0=R_0$. 
The phase profile is 
\begin{align}
 \Psi_{nm}(f)&=\arg \Gamma_{nm}(f)=-2\arctan \frac{X_{nm}(f)}{R_0}  \, .
\end{align}

Suppose the reactive impedance consists of a resonating series $LC$ circuit \cite{SelEle:13,Gly:16,HumCar:14,TayTanPalUdpUdpRot:15,MohNasAnd:16,Bur:16}, with cell-dependent inductive and capacitive values $L_{nm}$ and  $C_{nm}$, respectively. The corresponding impedance is 
\begin{align}
	Z_{nm}(f)=\jmath X_{nm}(f)=-\jmath \frac{1-(2\pi f)^2 L_{nm} C_{nm}}{2\pi f C_{nm}} \, ,
\end{align}
where $L_{nm}$ and  $C_{nm}$ are chosen to satisfy $(2\pi \sqrt{L_{nm} C_{nm} })^{-1}=\fr$.

To obtain the form in \eqref{eq:Psif}, it is convenient to derive the first-order Taylor series expression in $f$ for the reflection coefficient phase with respect to the reference frequency $\fr$. In particular, for the $LC$ load it results
\begin{align} \label{eq:PsiLC}
  \Psi_{nm}(f)& \approx -\frac{8 \pi L_{nm}}{R_0}  (f -\fr) \, .
\end{align}

From \eqref{eq:PsiLC} it is possible  to obtain the desired coefficient $\alpha_{nm}=-8 \pi L_{nm}/R_0$ in \eqref{eq:Psif} by properly designing $L_{nm}$ and  $C_{nm}$ in each cell.

\subsection{Reflections from the Environment}

A location in \ac{NLOS} channel condition might still be covered without introducing metaprisms if a sufficient \ac{SNR} is obtained from the signal scattered by the surrounding walls (e.g., buildings).
Therefore, for a fair comparison between the performance achieved with and without metaprisms, it is important to consider also the signals scattered by walls. 
To this purpose, in this section we  summarize the common models used to describe the scattering process in typical rough surfaces like walls.


As reference, we consider a typical scattering process which can be modeled as the superimposition of a specular component obeying the Snell's law and a diffuse scattering component. The latter can be further modeled according to the widely-used Lambertian model\cite{Deg:01,DegFusVitFal:07}. 

For convenience, we discretize the wall in the same way as the metaprism, with small areas $\Ac$ (cells) at  positions $\boldp_{nm}$, with $n=1,2,\ldots N_w-1$ and $m=1,2, \ldots M_w-1$.\footnote{$N_w$ and $M_w$ are in general larger than $N$ and $M$ as walls have typically a larger extension.} The Lambertian model can be equivalently described in terms of reflection coefficients of each cell, in the presence of an incident plane wave with 2D angle $\Thetai$ and observed at angle $\Theta$, as follows
\begin{equation} \label{eq:rnms}
r_{nm}(\Thetai,\Theta)=\Gamma(\thetai)\, R \, \sqrt{\Gs} \, + S \, \sqrt{\Gs \cos \thetai \, \cos \theta} \,e^{\jmath \Psi_{nm}(\Thetai,\Theta)} \, ,
\end{equation}
where $\Gs=\Ac \frac{4 \pi}{\lambda^2}$, $\Gamma(\thetai)$ is the Fresnel reflection coefficient, $R$ is the reflection reduction factor, and $S$ is the  scattering coefficient.   In \eqref{eq:rnms} the first term corresponds to the specular component (since all cells have the same phase, the reflection results to be specular), whereas the second component refers to the Lambertian scattering pattern.  
The phase shift $\Psi_{nm}(\Thetai,\Theta)$ at position $p_n$ is determined so that the  signal reflected by all cells sum up coherently towards angle $\Theta$ when analyzing the scattering coefficient in that direction, i.e.,  
\begin{align} 
\Psi_{nm}(\Thetai,\Theta)= -  \frac{2 \pi }{\lambda} 
\left (  n d_x\left (   \ux(\Thetai)  + \ux(\Theta) \right ) + m d_y \left ( \uy(\Thetai) +  \uy(\Theta)\right )   \right )  +\Psi_0  \, , 
\end{align}
where $\Psi_0$ is a common phase offset and, for convenience, we have defined the quantities $\ux(\Theta)=\sin(\theta) \cos(\phi)$ and $\uy(\Theta)=\sin(\theta) \sin(\phi)$.


The  Fresnel reflection coefficient $\Gamma(\thetai)$  can be found in many text books as a function of the refraction coefficient of the medium $n$ and the incident angle $\thetai$. For instance, for the \ac{TE} polarization it is \cite{LanFeuRap:96}
\begin{equation}
\Gamma(\thetai)=\frac{\cos \thetai - \sqrt{n^2 - \sin^2 \thetai}}{\cos \thetai + \sqrt{n^2 - \sin^2 \thetai}} \, ,
\end{equation}
where $n^2 = \epsilon_r - j \epsilon_r \tan \delta$, with $\epsilon_r$  being the relative dielectric constant, and $\tan \delta$ being the loss tangent.
For what the reflection reduction factor $R$ is regarded, it can be derived by applying the theory of scattering from rough surfaces as a function of the standard deviation of surface roughness  \cite{Ame:53}. Alternatively, one could  obtain $R$ and $S$ from scattered field measurements 
\cite{DegFusVitFal:07}. 

\begin{figure}[t]
\centerline{\includegraphics[width=0.7\columnwidth]{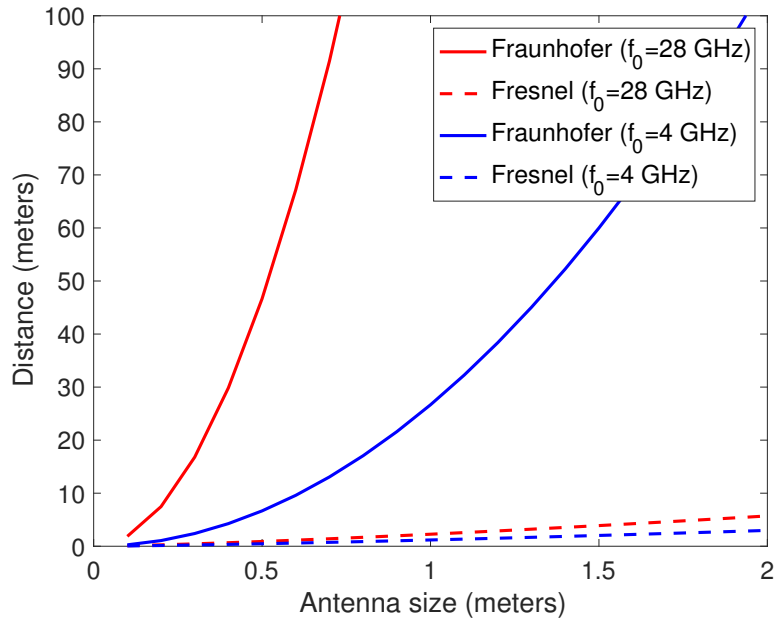}}
\caption{Fraunhofer and Fresnel distances as a function of antenna size $D$, for $f_0=4\,$GHz and $f_0=28\,$GHz.}
\label{Fig:FDistance}
\end{figure}  
  
\subsection{Near-field and Far-field Regions}  
  
For further convenience, it is worth to define the Fraunhofer and Fresnel distances  \cite{BalB:16}
\begin{align}
     \dfraun=\frac{2\,D^2}{\lambda} & \hspace{2cm}
    \dfresn=\sqrt[3]{\frac{D^4}{8\,\lambda}} \, ,
\end{align}
where $D=\max(\Lx,\Ly)$ is the antenna diameter.
At distance $d>\dfraun$ between the surface and the TX/RX antennas, the system is operating in the far-field region and only beamsteering is possible with antenna arrays or metasurfaces.
Instead when $d<\dfraun$, the system is operating in the near-field region\footnote{Here we are considering the radiating near-field region. The reactive near-field component becomes significant at distances typically less than $\dfresn$ \cite{BalB:16}.} and focusing is possible \cite{NepBuf:17}. 
In Fig. \ref{Fig:FDistance}, $\dfraun$ and $\dfresn$ are shown as a function of antenna diameter $D$ for two different frequency bands. As it can be noticed, when operating at millimeter waves, the far-field assumption becomes no longer valid  also at distances of dozen of meters for relative small antennas/surfaces. Therefore, design approaches based on this assumption should be revisited as done in this paper.

\section{System Model}
\label{Sec:SystemModel}

\subsection{Scenario Considered}

We consider a downlink \ac{OFDM}-based wireless system where a \ac{BS} serves  $U$ users located in \ac{NLOS} condition with respect to it, as illustrated in Fig. \ref{Fig:Scenario}. The communication between the \ac{BS} and the generic user might be possible in some locations thanks to the specular and diffuse scattering of the signal  from  the portion of the wall illuminated by the transmitted signal. Nevertheless, the communication performance can be significantly enhanced by deploying a frequency-selective metasurface, i.e.,  a metaprism, 
 as it will be evident in the numerical results. 
We assume the \ac{BS} is in  \ac{LOS} condition and in the far-field region with respect to the surface (metaprism or wall).  Moreover, we consider all users are located in  \ac{LOS} condition with respect to the surface but they could be in near- or far-field region depending on their distance from the surface.

In a conventional \ac{OFDM} system, the total bandwidth $W$ is equally divided into $K\ge U$ orthogonal subcarriers with subcarrier spacing $\Delta f=W/K$. The frequency of the $k$th subcarrier is $f_k=f_0-W/2+k\Delta f$, $k=1,2,\ldots , K$, where $f_0$ denotes the central  frequency.

To each user $u$, with $u \in \{1,2 , \ldots U \}$, a specific subcarrier $k=A(u)$ is one-to-one assigned according to some policy, as it will specified in Sec.\ref{Sec:AR},  where $A(u)$ is the assignment function and $A^{-1}(k)$ is its inverse.  The case where a user requires more subcarriers (higher traffic) can be easily managed by grouping different subcarriers into a single resource block.

Given user $u$, denote with $x^{(k)} \in {\cal{C}}$, with $\EX{\left (x^{(k)} \right )^2 }=1$ and $k=A(u)$, its  information symbol transmitted at the generic \ac{OFDM} frame.\footnote{$\EX{.}$ denotes the statistical expectation operator.}
The total transmit power $\Ptx$ is allocated differently among subcarriers (and hence users) by multiplying the corresponding transmitted symbols by the weights $\omega^{(k)}$, $k=1,2,\ldots , K$, which account for the relative power assigned to each subcarrier under the constrain $\sum_k \left ( \omega^{(k)} \right )^2=1$.



Consider the impinging plane wave emitted by the transmit \ac{BS} located in position $\boldpbs$ with \ac{AOA} $\Thetai=(\thetai, \phii)$, with respect to the normal direction of the metaprism, and distance $|\boldpbs-\boldp_0|=|\boldpbs|$.

When the relative bandwidth satisfies $W/f_0\ll1$, the (complex) channel gain between the transmitter and the $nm$th cell of the metaprism for the $k$th subcarrier is 
\begin{align} \label{eq:h1}
h_{nm}^{(k)}(\boldpbs) &= \frac{\sqrt{\Gt} \lambda}{4\pi |\boldpbs - \boldp_{nm|}|} \exp \left (-\jmath \frac{2 \pi f_k}{c}  |\boldpbs-\boldp_{nm}| \right )   \, ,
\end{align}
where $\Gt$ is the transmit antenna gain, and $\lambda=c/f_0$, with  $c$ being the speed of light. 

In far-field condition, \eqref{eq:h1} can be well approximated as
\begin{align} \label{eq:h}
h_{nm}^{(k)}(\boldpbs) & \simeq  \frac{h_0}{|\boldpbs|}  \exp \left (\jmath \frac{2 \pi}{\lambda}   \left ( n d_x  \, \ux(\Thetai)  +m d_y \, \uy(\Thetai) \right )    \right ) \, ,
\end{align}
where $h_0=\sqrt{\Gt} \frac{ \lambda}{4\pi} \exp \left ( -\jmath \frac{2 \pi}{\lambda} |\boldpbs|  \right )$.
The exponential argument accounts for the phase shift with respect to the metaprism center $\boldp_0$.
 
Similarly, the  channel gain  from  the $nm$th cell to the receiver located in position $\boldp=(x,y,z)$ (reflection channel)  is 
\begin{align} \label{eq:gk}
g_{nm}^{(k)}(\boldp)= \frac{\sqrt{\Gr} \lambda}{4 \pi |\boldp-\boldp_{nm}|} \exp \left (- \jmath \frac{2\pi f_k}{c} |\boldp-\boldp_{nm}| \right )    \, ,
\end{align}
where $\Gr$ is the receive antenna gain. 
%
In the far-field region, 
the previous expression can be approximated as
\begin{align}\label{eq:gkapprox}
g_{nm}^{(k)}(\boldp)\simeq \frac{g_0}{|\boldp|}  \exp \left (\jmath \frac{2 \pi}{\lambda}   \left ( n d_x  \, \ux(\Theta)  +m d_y \, \uy(\Theta) \right )    \right ) \, ,
\end{align}
where $\Theta=(\theta,\phi)$ is the angle corresponding to position $\boldp$, and
$g_0  =\sqrt{\Gr} \frac{\lambda}{4\pi} \exp \left ( -\jmath \frac{2 \pi}{\lambda} |\boldp|  \right )$.  

By combining the previous expressions, the received signal at the $k$-th subcarrier  is given by 
\begin{align} \label{eq:yk}
y^{(k)}&= \sum_{n=0}^{N-1} \sum_{m=0}^{M-1} h_{nm}^{(k)} (\boldpbs) \, r_{nm}^{(k)}(\Thetai,\Theta) \, g_{nm}^{(k)} (\boldp) \, \sqrt{\Ptx}\,  \omega^{(k)} \, x^{(k)} + n^{(k)} \nonumber \\
& =\sqrt{\Ptx}\, c^{(k)}(\boldpbs,\boldp) \, \omega^{(k)} \, x^{(k)}+ n^{(k)} \, ,
\end{align}
where  
$n^{(k)}$ is the thermal noise modeled as a zero-mean complex circular symmetric Gaussian \ac{RV} with variance $\sigma_n^2$,
and  $r_{nm}^{(k)}(\Thetai,\Theta)=r_{nm}(\Thetai,\Theta;f_k)$ is the surface reflection coefficient at frequency $f_k$ given by  \eqref{eq:rnm1}.
The phase shift and gain undertaken by the $k$th subcarrier from the $nm$th cell are, respectively,
\begin{align}
\Psi_{nm}^{(k)}=\Psi_{nm}(f_k)  & \hspace{1cm} \beta_{nm}^{(k)}(\Thetai,\Theta)=\beta_{nm}(\Thetai,\Theta;f_k) \, .
\end{align}

It is worth to notice that the detection of $x^{(k)}$ requires only the estimation of the global channel coefficient $c^{(k)}(\boldpbs,\boldp)$, i.e., the end-to-end \ac{CSI}, which includes the \ac{BS}-metaprism and metaprism-user channels. Instead, with a \ac{RIS} one has to estimate them singularly thus making the \ac{CSI} a challenging task.

\section{Subcarrier-dependent Beamsteering}
\label{Sec:Beamsteering}


In this section, we investigate how, through a proper design of the metaprism coefficients $\alpha_{nm}$ in \eqref{eq:Psif},   it is possible to perform subcarrier-dependent beamsteering. This requires that both the transmitter and the receiver are in the \ac{LOS} far-field region with respect to the metaprism. 
We consider the particular but significant case where $ \beta_{nm}^{(k)}(\Thetai,\Theta)=\beta_0^{(k)}(\Thetai,\Theta)$, $\forall n,m$, 
so that using the approximations \eqref{eq:h} and \eqref{eq:gkapprox},  equation \eqref{eq:yk} can be written as 
\begin{align} \label{eq:yk1}
 y^{(k)}=& 
 \frac{ \sqrt{\Ptx}\,  h_0\,  g_0 \, \beta_0^{(k)}(\Thetai,\Theta)\,  }{|\boldpbs|\, |\boldp|} \, \omega^{(k)}\,  x^{(k)}
   \,   \nonumber \\
& \cdot \sum_{n=0}^{N-1} \sum_{m=0}^{M-1} \exp \left (\jmath \frac{2 \pi }{\lambda} 
\left (  n d_x\left (  \ux(\Thetai)  + \ux(\Theta) \right ) + m d_y \left (\uy(\Thetai)  +\, \uy(\Theta) \right )   \right ) + \jmath \, \Psi_{nm}^{(k)}    \right )  + n^{(k)} \, .
\end{align}

For instance, supposing a perfect specular reflector, i.e., $\Psi_{nm}^{(k)}=\pi$, $\beta_0^{(k)}(\Thetai,\Theta)=1$, 
$\forall n,m$, and $\phii=\phi=0$, the maximum channel gain is obtained when $\theta=-\thetai$ (Snell's law).\footnote{To lighten the treatment, sometimes we take  the liberty of derogating from the conventional spherical coordinate system by allowing $\theta$ ranging in $[-\pi/2, \pi/2)$ and $\phi \in [0,\pi)$.     }  
From \eqref{eq:yk1}, the signal received by a user located at angle  $\theta=-\thetai$  simplifies to  
\begin{equation}
    y^{(k)}=\jmath \frac{\sqrt{\Ptx}\, h_0\,    g_0 \,   \ }{|\boldpbs|\, |\boldp|}\, \omega^{(k)}\,  x^{(k)} + n^{(k)} \, .
\end{equation}

More in  general, to point the beam towards a target direction $\Theta_0^{(k)}=\left (\theta_0^{(k)},\phi_0^{(k)} \right )$ for the $k$th subcarrier, the phase profile should be designed such that 
\begin{align} \label{eq:Psibeamforming}
\Psi_{nm}^{(k)}=  - \frac{2 \pi }{\lambda} 
\left (  n d_x\left (   \ux(\Thetai)  + \ux \left (\Theta_0^{(k)} \right ) \right ) + m d_y \left ( \uy(\Thetai) +  \uy \left (\Theta_0^{(k)} \right )\right )   \right )    \, , 
\end{align}
so that all phasors in \eqref{eq:yk1} sum up coherently towards the direction $\Theta_0^{(k)}$.

Note that one could see the system transmitter+metaprism as an equivalent planar antenna array (reflectenna)  whose frequency-dependent array factor is 
\begin{align} \label{eq:AF}
AF^{(k)}(\Theta)= 
& \sum_{n=0}^{N-1} \sum_{m=0}^{M-1} \exp \left (
  \jmath \frac{2 \pi n d_x}{\lambda}  \left (  \ux(\Theta)  - \ux \left (\Theta_0^{(k)} \right ) \right ) +  \jmath \frac{2 \pi m d_y}{\lambda} \left (\uy(\Theta)  -\, \uy \left ( \Theta_0^{(k)} \right ) \right )      \right )   \, ,
\end{align}
having $\beta_0^{(k)}(\Thetai,\Theta)$ as single-element antenna pattern.

Consider now the reflecting  metaprism has been designed such that 
the following frequency-dependent phase profile  (beamsteering phase profile) holds 
\begin{align} \label{eq:psi_far}
\Psi_{nm} (f)= \alpha_{nm} \cdot (f-f_0) =  ( a_0\,  x_n + b_0\,   y_m ) \cdot (f-f_0) \, ,
\end{align}
by setting $\fr=f_0$  in \eqref{eq:Psif}, with $a_0$ and $b_0$ being properly chosen constants.

By equating \eqref{eq:psi_far} and \eqref{eq:Psibeamforming} it is 
\begin{align} \label{eq:a0b0gen}
a_0 (f_k -f_0)& = - \frac{2 \pi  }{\lambda} \left (   \ux(\Thetai)  +   \ux\left (\Theta_0^{(k)}\right )  \right )     \nonumber \\
b_0 (f_k -f_0)&= - \frac{2 \pi  }{\lambda} \left ( \uy(\Thetai) +  \uy\left (\Theta_0^{(k)} \right ) \right )    \, ,
\end{align} 
from which we can determine the reflection direction $\Theta_0^{(k)}$  as a function of the subcarrier  $k$
\begin{align} \label{eq:thetak}
\ux\left (\Theta_0^{(k)}\right ) =& -\ux(\Thetai) - \frac{a_0\, \lambda}{2\pi}(f_k -f_0) \nonumber \\
\uy\left (\Theta_0^{(k)}\right ) =& -\uy(\Thetai) - \frac{b_0\, \lambda}{2\pi}(f_k -f_0)  \, ,
\end{align} 
which can be arranged as
\begin{align} \label{eq:uxuy}
\phi_0^{(k)}&=\arctan \frac{\uy(\Thetai) + b_0\, \lambda\, (f_k -f_0)/ 2\pi }{\ux(\Thetai) + a_0\, \lambda \, (f_k -f_0) /2\pi} \nonumber \\
\theta_0^{(k)}&=-\arcsin \frac{\ux(\Thetai) + a_0\, \lambda\, (f_k -f_0)/2\pi}{\cos \phi_0^{(k)}}
 \, .
\end{align} 


\begin{figure}[t]
\centering
\subfigure[][ $\thetai=45^{\circ}\,$, $\phii=0^{\circ}$, $\theta_m=40^{\circ}$.]{\label{Fig:AF}\includegraphics[width=0.5\textwidth]{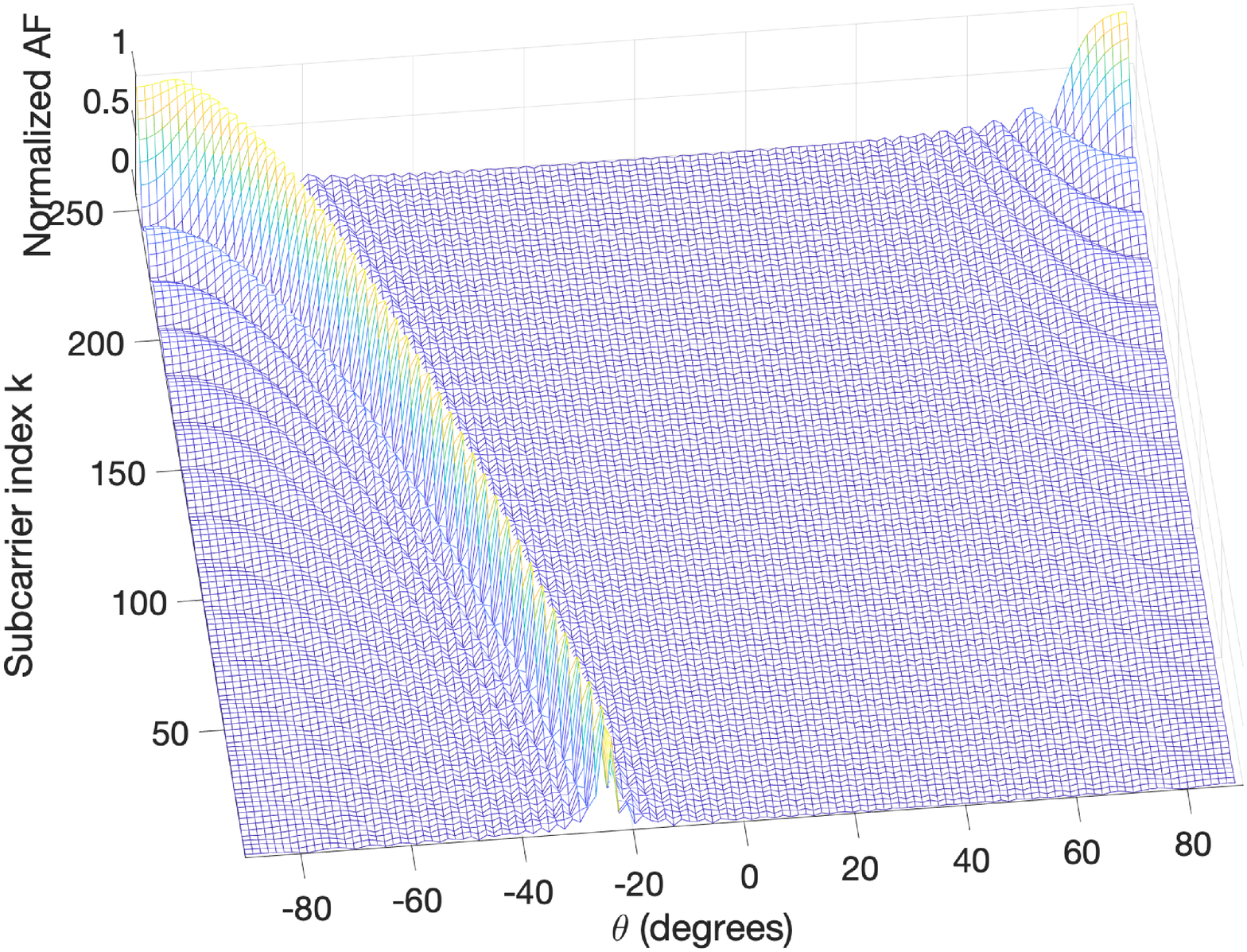}}
\subfigure[][$\thetai=0^{\circ}\,$, $\phii=0^{\circ}$, $\theta_m=90^{\circ}$.]{\label{Fig:AF2}\includegraphics[width=0.45\textwidth]{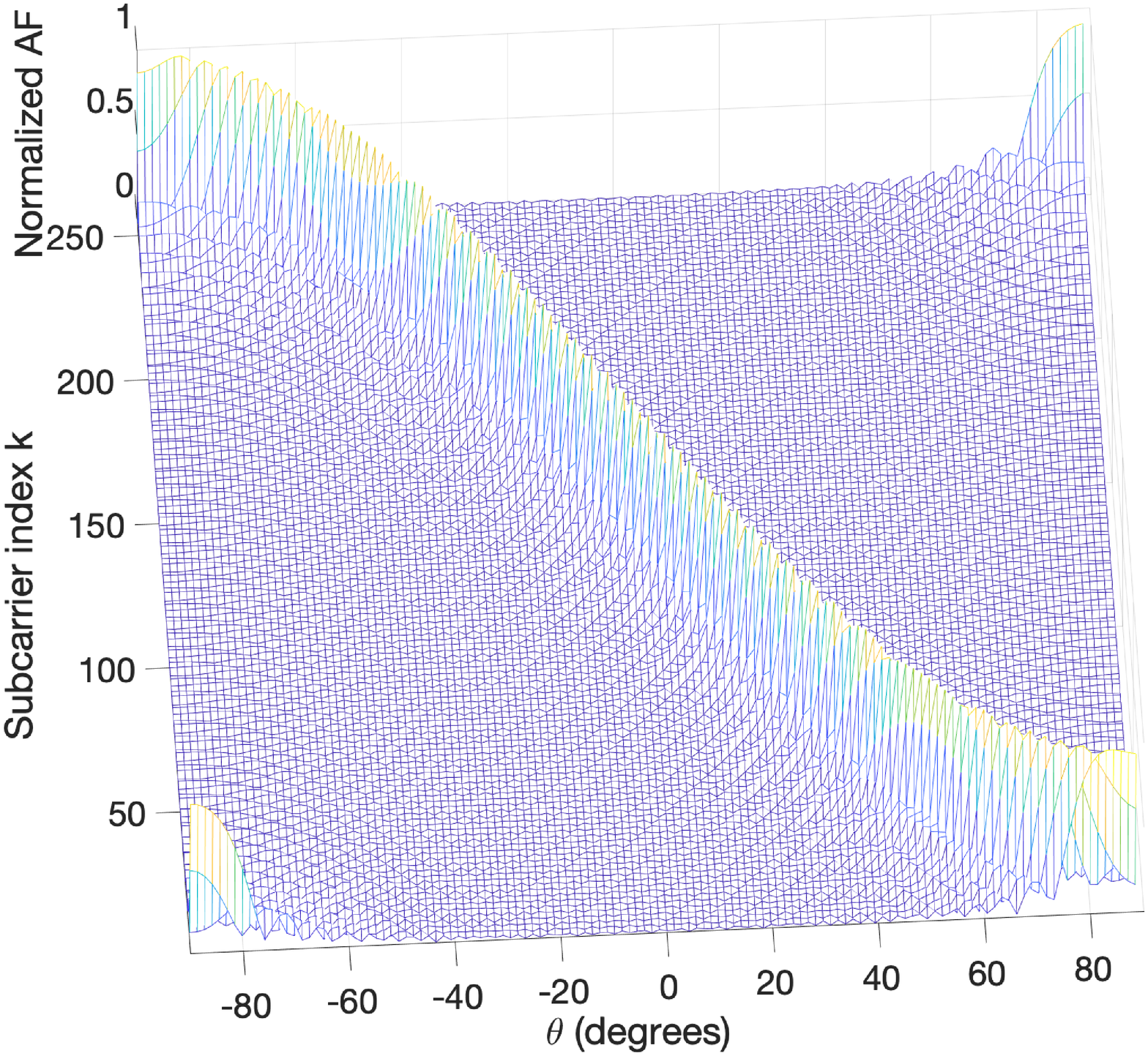}}
\caption{Normalized equivalent array factor as a function of the subcarrier index $k$. $N=M=50$, $d_x=d_y=\lambda/2$.  }
\end{figure}

\subsection{Design Example}
\label{Sec:DesignBeamsteering}

Suppose one wants the metaprism reflects  an impinging signal, with incident angle $\Thetai=(\thetai,0)$, in the $x-z$ plane  dependently on the subcarrier index $k$. 

For the central subcarrier $k_0=K/2$, corresponding to $f_{k_0}=f_0$, according to \eqref{eq:uxuy} it is 
$\theta_0^{(k_0)}=-\thetai$, independently of $a_0$ and $b_0$.
For instance, coefficients $a_0$ and $b_0$ can be designed so that when $k=K$ (highest subcarrier) it is  $\theta_0^{(K)}=-\theta_i-\theta_m$, for some angle $\theta_m$. From \eqref{eq:a0b0gen} by setting $k=K$ it is 
\begin{align} \label{eq:a0b0}
a_0 & = - \frac{2 \pi }{\lambda (f_K -f_0)} \left (   \ux(\Thetai)  +   \ux \left (\Theta_0^{(K)} \right )  \right ) = -\frac{4 \pi }{\lambda W} \left (- \sin (\thetai +  \theta_m)  + \sin \thetai     \right )    \nonumber \\
b_0 &= - \frac{2 \pi }{\lambda (f_K -f_0)} \left ( \uy(\Thetai) + \, \uy \left (\Theta_0^{(K)} \right ) \right )   =0 \, .
\end{align} 

With this choice of $a_0$ and $b_0$, from \eqref{eq:thetak} we obtain
\begin{equation}
	\sin \left (\theta_0^{(k)}\right ) = -\sin(\thetai) + \frac{2(f_k-f_0)}{W} \left ( \sin(-\thetai-\theta_m) + \sin(\thetai)  \right ) \, .
\end{equation}

The extreme case when $k=1$ (lowest subcarrier) gives $\sin\left (\theta_0^{(1)} \right )=-2 \sin (\thetai)+ \sin(\theta_i+\theta_m$).
As result, each subcarrier is reflected according to a different angle in the range $\left [\theta_0^{(K)},\theta_0^{(1)} \right]$, thus creating the ``prism" behavior.

A numerical example is provided in Fig. \ref{Fig:AF}, where the  equivalent array factor given by \eqref{eq:AF}, normalized with respect to $N\times M$,  is shown for each subcarrier. Coefficients $a_0$ and $b_0$ have been computed using \eqref{eq:a0b0} with the following parameters: 
$\thetai=45^{\circ}$, $\phii=0^{\circ}$, $\theta_m=40^{\circ}$, $\lambda \, W=10^6$ and $K=256$. 
It can be easily verified that the main lobe of the radiation pattern shifts from $\theta_0^{(1)}\simeq -25^{\circ}$  to $\theta_0^{(K)}=-\theta_i-\theta_m=-85^{\circ}$, when the subcarrier index ranges from 1 to $K$.

Another example is given in Fig. \ref{Fig:AF2} where $\thetai=0^{\circ}$, $\phii=0^{\circ}$, $\theta_m=90^{\circ}$, corresponding to coefficient $a_0=-4 \pi /\lambda W$. With this design, the main lobe of the radiation pattern spans in the range $[-90^{\circ}, -90^{\circ} ]$.

\section{Subcarrier-dependent Focusing}
\label{Sec:DesignFocusing} 

In the following, we suppose the receiver is in the near-field region with respect to the metaprism whereas the \ac{BS} is still in the far-field.
Supposing an incident signal with \ac{AOA} $\Thetai$, if one wants to focus the signal on position $\boldp$, with angle $\Theta_0^{(k)}$ and distance $d_{\text{F}}^{(k)}=|\boldp|$ so that $\dfresn<d_{\text{F}}^{(k)}<\dfraun$, all components in \eqref{eq:yk} must sum up coherently at that position. Considering the following approximation   $|\boldp-\boldp_{nm}|\simeq |\boldp|$, which holds when the receiver is not too close to the metaprism, this corresponds to designing the metaprism with the following  phase profile for the $k$th subcarrier  (Fresnel approximation) \cite{NepBuf:17}
\begin{align}  \label{eq:psifocus}
\Psi_{nm}^{(k)}= \frac{2\pi}{\lambda} \frac{\left (x_n^2+ y_m^2 \right )}{2 d_{\text{F}}^{(k)}}   -  \frac{2 \pi }{\lambda} 
\left (  x_n\left (   \ux(\Thetai)  + \ux \left (\Theta_0^{(k)} \right ) \right ) + y_m \left ( \uy(\Thetai) +  \uy \left (\Theta_0^{(k)} \right )\right )   \right )  \, .
\end{align}

Note that now in \eqref{eq:yk}, the exact expression \eqref{eq:gk} should be used instead of \eqref{eq:gkapprox}.

It is convenient to choose $\fr=f_1$ in \eqref{eq:Psif}  so that $d_{\text{F}}^{(k)}>0$, $\forall k $. It follows from \eqref{eq:Psif} and \eqref{eq:psifocus} that  
\begin{align} \label{eq:Focusing}
\Psi_{nm}^{(k)}= \alpha_{nm} \cdot (f_k-f_1) = \left [ a_{\text{F}}\,  \left (   x_n^2 +  y_m^2 \right ) + a_0\, x_n + b_0\, y_m \right ] \cdot (f_k-f_1) \, ,
\end{align}
from which one can determine the focal distance obtained at each frequency
\begin{align} 
d_{\text{F}}^{(k)}= \frac{\pi f_1}{c\,  a_{\text{F}}\,  (f_k-f_1)}\, ,
\end{align}
as a function of parameter $a_{\text{F}}$, whereas $a_0$ and $b_0$ can be designed according to the criteria given in the previous section once the target direction $\Theta_0^{(k)}$ is fixed.

\subsection{Design Example}

For instance, parameter $a_{\text{F}}$ could be designed so that, given a minimum desired focal distance $d_m$, 
$d_{\text{F}}^{(k)}=d_m$ when $k=K$, i.e., 
\begin{align} 
a_{\text{F}}=  \frac{2\pi f_1}{c\,  d_m\,  W}\, .
\end{align}

When moving to lower subcarrier indexes the focal distance will increase to infinity (when $k=1$).
It is worth to notice that the phase profile in \eqref{eq:Focusing} degenerates to that of beamsteering in \eqref{eq:Psibeamforming} (with $f_1$ instead of $f_0$) when approaching the far-field region, i.e., when $d_{\text{F}}^{(k)} \approx \dfraun$, or entering the far-field region. This means that when increasing the focal distance it is no longer possible to discriminate distances but only angles. 
Focusing can be helpful when one is interested in discriminating users located at different distances but with similar angles as happens in mono-dimensional scenarios such as along corridors or streets. An example will be presented in the numerical results.

%
%
%
%

\section{Some Considerations about Path-Loss using Metasurfaces}
\label{Sec:PL}

To compute the link-budget using metaprisms and, in general, metasurfaces  in \acp{RIS}, it is important to understand how  the total path-loss depends on system parameters, in particular on metasurface dimensions and characteristics. 
This topic has received some attention in recent literature regarding \ac{RIS}-enabled wireless networks \cite{TanCheDaiHanDiRZenJinCheCui:19,Ell:19,BjoSan:19,BjoSan:20}. However, in many cases the validity range of the models obtained are not properly investigated.

In general, for a given metaprism design, the path-loss for subcarrier $k$ can be computed from \eqref{eq:rnm1} and \eqref{eq:yk}
\begin{align} \label{eq:Lk}
L^{(k)}=& \left | c^{(k)}(\boldpbs,\boldp)    \right |^{-2}=\frac{(4 \pi)^4 }{\lambda^4 \Gt \Gr \Gc^2 \, F(\Thetai) \, F(\Theta)  }    \nonumber \\
& \cdot \left | \sum_{n=0}^{N-1} \sum_{m=0}^{M-1} \frac{\Gamma_{nm}(f)}{|\boldpbs - \boldp_{nm|}|\, |\boldp-\boldp_{nm}|} \exp \left (-\jmath \frac{2 \pi f_k}{c} \left ( |\boldpbs-\boldp_{nm}| -  |\boldp-\boldp_{nm}| \right ) \right )
   \right |^{-2} \, .
\end{align} 

The path-loss can be minimized by properly designing the amplitude and phase profiles of  $\Gamma_{nm}(f)$ under the constraint that $|\Gamma_{nm}(f)|\le 1$ since the metaprism is supposed to be passive. 

In the particular but significant case where $|\Gamma_{nm}(f)|=1$,  $|\boldpbs - \boldp_{nm|}|\simeq |\boldpbs|$, $|\boldp - \boldp_{nm|}|\simeq |\boldp|$ (amplitude approximation, i.e., transmitter and receiver not too close to the metaprism) and the  phase profile $\Psi_{nm}(k)$ of the metaprism is designed to perfectly compensate the phase distortion introduced by the channel, i.e., 
\begin{align} \label{eq:IdealPhase}
\Psi_{nm}^{(k)}=\frac{2 \pi f_k}{c} \left ( |\boldpbs-\boldp_{nm}| +  |\boldp-\boldp_{nm}| \right ) \, ,
\end{align} 
eq. \eqref{eq:Lk} simplifies as 
%
\begin{align} \label{eq:Lkapprox}
L^{(k)} \simeq &   \frac{(4 \pi)^4 |\boldpbs|^2\, |\boldp|^2 }{\lambda^4 \Gt \Gr \Gc^2 \, F(\Thetai) \, F(\Theta)\, (N\, M)^2  }     
\, ,
\end{align} 
which was obtained also by other authors \cite{TanCheDaiHanDiRZenJinCheCui:19,Ell:19,BjoSan:19} under far-field conditions for \acp{RIS}.
Actually, the path-loss in \eqref{eq:Lkapprox} is valid also in near-field conditions provided the perfect phase profile \eqref{eq:IdealPhase} is considered. Unfortunately, its implementation might be challenging from the metaprism design and practical points of view, as it requires a very accurate information about user's position. Therefore, simplified but approximate phase profile design methods are preferred, such as beamsteering and focusing.  

Equation \eqref{eq:Lkapprox} may give the false illusion that by augmenting the number $N\times M$ of cells of the metasurface, the path-loss could be reduced arbitrarily. Unfortunately, this is true only under certain conditions and one should be careful in extrapolating this result, as commented in the following 3 arguments.

\begin{figure}[t]
\centering
\subfigure[][ $L_x=L_y=25\,$cm.]{\label{Fig:PL1}\includegraphics[width=0.48\textwidth]{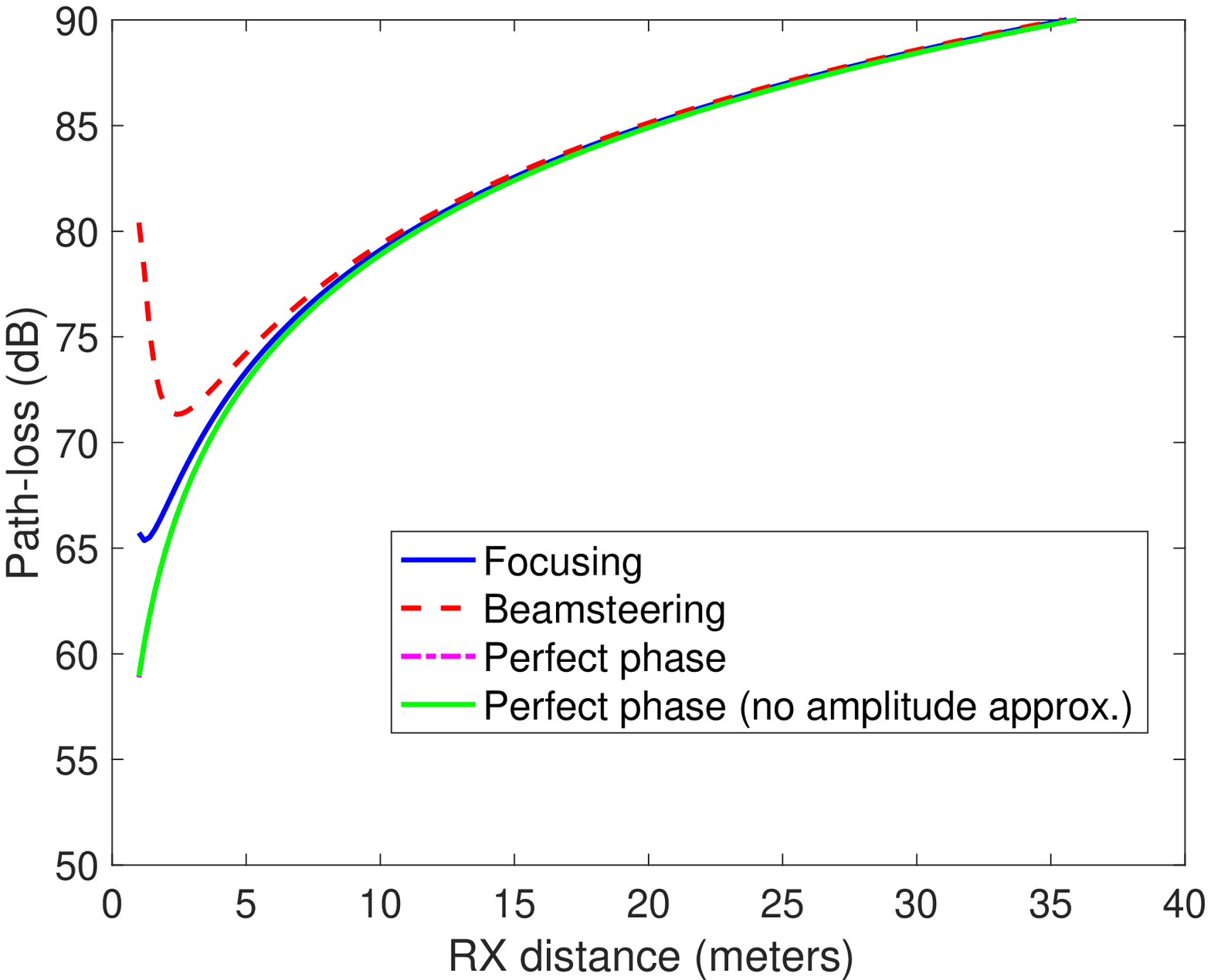}}
\subfigure[][$L_x=L_y=50\,$cm.]{\label{Fig:PL2}\includegraphics[width=0.48\textwidth]{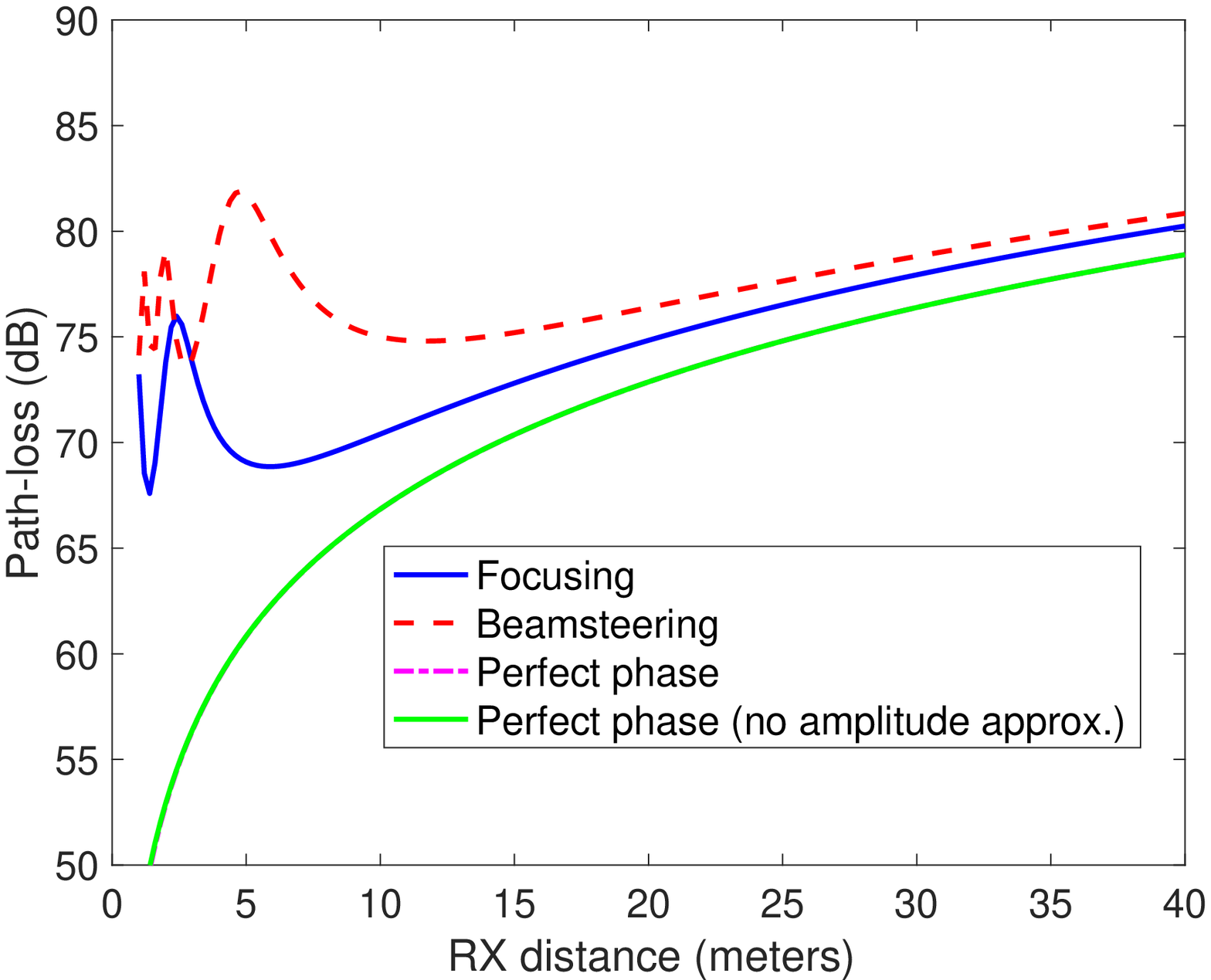}}
\caption{Path-loss as a function of RX distance. TX located at 20 meters. Both TX and RX are located along the metaprism boresigth direction, $d_x=d_y=\lambda/2$, $f_0=28\,$Ghz, $\Gt=10\,$dB, $\Gr=2\,$dB. }
\end{figure}

First, if the phase profile of the metaprism has been designed to perform beamsteering (as in most of papers), \eqref{eq:Lkapprox}  is accurate only in far-field, i.e., when $|\boldp|$ and $|\boldpbs|$ approach or overcome $\dfraun$. When decreasing the distance, the actual path-loss could degrade with respect to that predicted by \eqref{eq:Lkapprox}. This is evident in the example shown in  Figs. \ref{Fig:PL1} and \ref{Fig:PL2}, related to two different dimensions of the metaprism, where the path-loss obtained using \eqref{eq:Lk} with phase profiles \eqref{eq:Psibeamforming} (beamsteering), \eqref{eq:psifocus} (focusing), and 
\eqref{eq:IdealPhase} (i.e., using \eqref{eq:Lkapprox}) as a function of receiver distance is reported. At short distances,   \eqref{eq:Lkapprox} could lead to very optimistic predictions if the phase profile has been designed using beamsteering, especially when large metaprisms are deployed. 
More favorable path-loss values can be obtained if one considers the focusing phase profile \eqref{eq:psifocus}.
Approaching the Fresnel distance $\dfresn < 50\,$cm, also focusing using \eqref{eq:psifocus}  becomes inaccurate and one has to consider the exact phase profile in \eqref{eq:IdealPhase}, but  near-field reactive effects will also emerge which make the analysis impossible without resorting to dedicated modeling at \ac{e.m.} level.
From  Fig. \ref{Fig:PL2}, it is also evident that the distance approximation is in general accurate even at very short distances (magenta and green curves are almost overlapped). 

Second, expression \eqref{eq:Lkapprox}, but also the general expression \eqref{eq:Lk}, assume implicitly that the whole metaprism is illuminated by the transmitted signal (for reciprocity, the same arguments hold also for the receiver). This could not be true because of shadowing caused by the obstacle (for instance, in Fig. \ref{Fig:Scenario} if one extended the metaprism towards the \ac{NLOS} area, part of the metaprism would not be illuminated by the \ac{BS}). Even in \ac{LOS} condition, only part of the metaprism will be illuminated if the transmitter (receiver) is  close to the metaprism and/or its antenna has a large gain. In fact, the higher is the antenna gain, the narrower is the illuminating beam.

Third, in the presence of large metaprisms in relation to the distance,  polarization mismatch could play an important role. Even if the transmit/receive antenna  and cell elements are designed and deployed with the same polarization (e.g., vertical), when one of the antennas is located  at a distance of the same order of magnitude as the metaprism's size, the cells at the edge of the metaprism might not be aligned with the impinging wave, thus generating a polarization mismatch which is not accounted for by  \eqref{eq:Lk}. A detailed analysis of the path-loss and the coupling modes when using \ac{LIS} antennas can be found in  \cite{Dar:19}.    

Before moving to the next section, it may be interesting to observe that when $\Thetai=\Theta=(0,0)$, \eqref{eq:Lkapprox} is nothing else than the radar equation in case of a perfect metal square  plate with area $A=L_x \cdot L_y$, whose \ac{RCS} at its boresight is $\rho_{\text{m}}=4\pi A^2 / \lambda^2$. In fact, it is $\Gc^2 \, F(\Thetai) \, F(\Theta)\, (N\, M)^2=(d_x d_y)^2 (N\, M)^2 (4\pi)^2 /\lambda^4=A^2 (4\pi)^2 /\lambda^4=\rho_{\text{m}} \, 4\pi /\lambda^2$, which inserted in \eqref{eq:Lkapprox} gives the well-known radar equation \cite{BalB:16}. At different angles, the equivalent directional \ac{RCS} of the metaprism is $\rho(\Thetai,\Theta)=4\pi A^2 F(\Thetai) \, F(\Theta)/\lambda^2$ (assuming the perfect phase profile \eqref{eq:IdealPhase}), where  the radiation pattern $F(\Theta)$  accounts also  for the fact that the effective area of the metaprism reduces when illuminated/observed at different angles. 
Summarizing, with an ideal design of the phase profile, the metaprism acts approximatively as a  perfect metal plate reflecting at the location of interest, while with a different design one has to be careful when using \eqref{eq:Lkapprox}.

\section{Achievable Rate}
\label{Sec:AR}

In the considered network shown in Fig. \ref{Fig:Scenario}, where $U$ users are served by one \ac{BS}, 
 the achievable data rate (bit/s/Hz) at user $u$ is given by
\begin{equation} \label{eq:Ru}
	R_u= \log_2(1+\SNR_u  ) \, ,
\end{equation}
being 
\begin{equation} \label{eq:SNR}
\SNR_u=\Ptx\, |c^{(A(u))}(\boldpbs,\boldp_u) \, \omega^{(A(u))}|^2/\sigma_n^2 \, ,
\end{equation}
 where $A(u)$ gives the subcarrier index assigned to user $u$ and $\boldp_u$ is the position of user $u$.  We assume $c^{(A(u))}(\boldpbs,\boldp_u)$ is perfectly estimated at the receiver (perfect total \ac{CSI}). 

During the  initial access, or periodically, each user in turn transmits a series of pilot symbols on all subcarriers and the \ac{BS} estimates the corresponding \acp{SNR}. Denote with $SNR(u,k)$, $u=1,2,\ldots U$, $k=1,2, \ldots K$, the \ac{SNR} measured by the \ac{BS} on subcarrier $k$ when user $u$ was accessing.
%
The problem is how to assign subcarriers to the $U$ users and how to determine the weights $\omega^{(k)}$, $k=1,2,\ldots K$, such that the per-user or network achievable rate is maximized. 

The general optimization appears prohibitive from the complexity point of view and it is out of the scope of this paper.
To obtain numerical results, we consider the  sub-optimal assignment Algorithm  \ref{Algorithm}, which aims at assigning subcarriers and weights such that all users experience the same achievable rate.
Other optimization strategies are possible depending on application requirements, which can be the topic of future works.

\begin{algorithm}[th]
\SetAlgoLined
\KwInput{Number of users $U$, SNR matrix $SNR(u,k)$}
\KwResult{Subcarriers assignment $\{A(u)\}$ and weights $\{ \omega(k) \}$ }
 $u=U$, $\omega(:)=0$\; 
 \While{$u > 0$}
 {
  $[msnr,iu,ik]=max(SNR(u,k))$ \tcp*{ search the couple user-subcarrier $(iu,ik)$ experiencing the max SNR $msnr$}
  \If{u==U}{
   $refsnr=msnr$ \tcp*{ use the absolute max SNR as reference}
   }
  $A(iu)=ik$ \tcp*{ assign subcarrier $ik$ to user $iu$ }
  $SNR(iu,:)=0$ \tcp*{ no longer consider user $iu$ and subcarrier $ik$}
  $SNR(:,ik)=0$ \tcp*{  in the next steps}
  $\omega(ik)=refsnr/msnr$ \tcp*{ compute the weights to equalize all SNRs}
  u=u-1 \;
 }
 $s=\text{sum}(\omega(k))$  \tcp*{ normalize the weights so that the sum is one}
 \For{u=1:U}{
 $\omega(A(u))=\omega(A(u))/s$ \;
 }
 \caption{Subcarrier and weights assignment.}
 \label{Algorithm}
\end{algorithm}

\begin{figure}[th]
\centerline{\includegraphics[width=0.7\columnwidth]{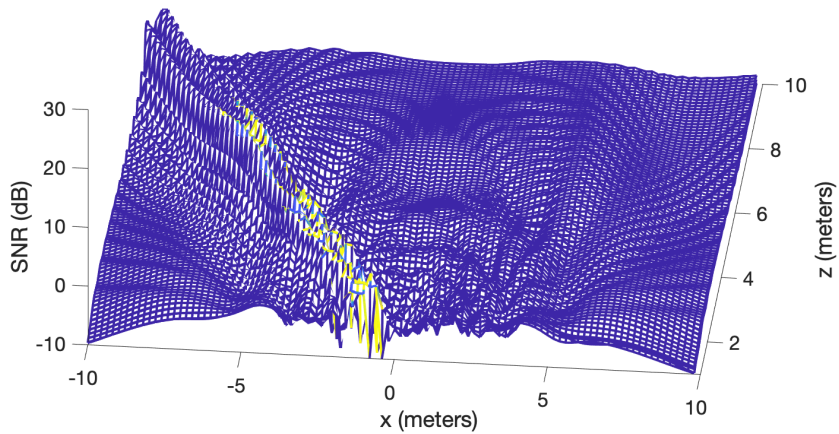}} 
\caption{SNR map due to scattering from the wall (absence of metaprism). Aerial concrete wall with illuminated area of $4\,\text{m}^2$.}
\label{Fig:SNRNoMeta}
\end{figure}

\begin{figure}[th]
\centerline{\includegraphics[width=0.7\columnwidth]{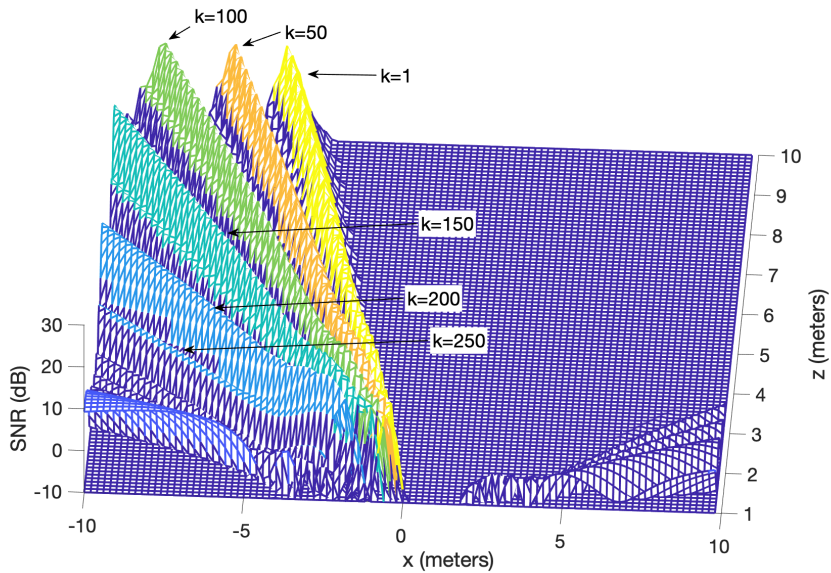}}  
\caption{SNR map with the metaprism. Beamsteering phase profile.  $L_x=L_y=50\,$cm. }
\label{Fig:SNRMeta}
\end{figure}

\begin{figure}[th]
\centerline{\includegraphics[width=0.7\columnwidth]{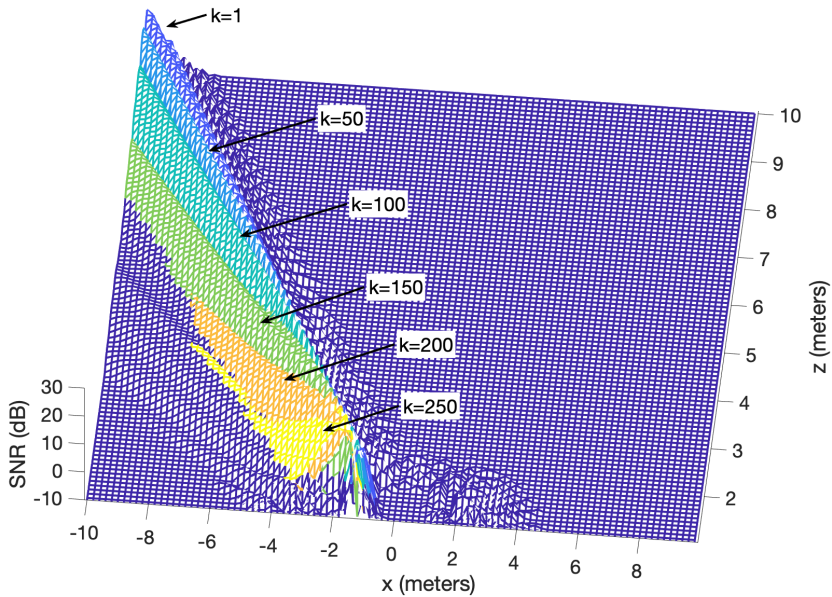}}  
\caption{SNR map with the metaprism. Focusing phase profile.  $L_x=L_y=50\,$cm. $d_m=2\,$meters, $\theta_m=5^{\circ}$. }
\label{Fig:SNRFocusing}
\end{figure}

\begin{figure}[th]
\centerline{\includegraphics[width=0.7\columnwidth]{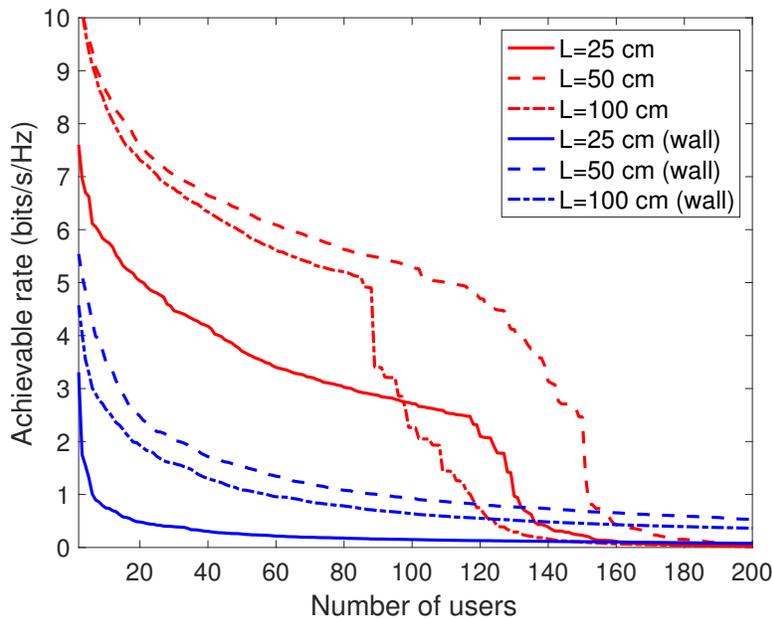}}
\caption{Per-user achievable rate vs number of users for different metaprism sizes.  }
\label{Fig:mAR}
\end{figure}

\begin{figure}[th]
\centerline{\includegraphics[width=0.7\columnwidth]{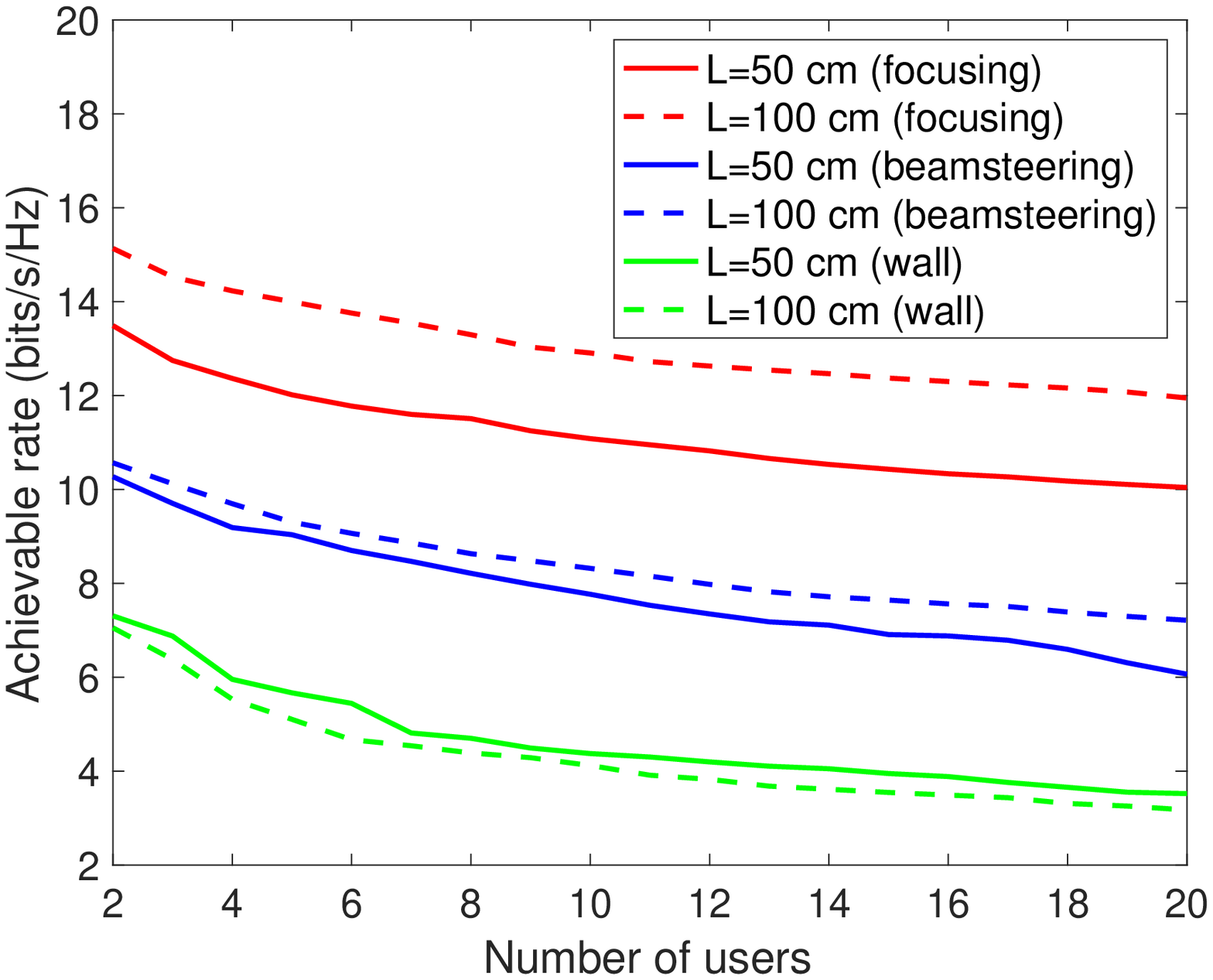}}
\caption{Per-user achievable rate vs number of users for different metaprism sizes. Mono-dimensional scenario.  }
\label{Fig:ARMono}
\end{figure}

\section{Numerical results}
\label{Sec:NumericalResults}

In the following, we corroborate the proposed metaprism idea with some numerical examples.

The scenario considered is that illustrated in  Fig. \ref{Fig:Scenario}, where a transmitting \ac{BS} is located at position $\boldpbs=(14.21,h,14.21)\,$meters, corresponding to a distance $d=20\,$meters from the metaprism and angle $\Thetai=(45^{\circ},0)$. We suppose the \ac{BS}, the metaprism and the users are approximatively at the same height so that, without loss of generality, we set $h=0$ (2D scenario).

If not otherwise specified, the following parameters have been used for  the transmitter and the receiver: $f_0=28\,$GHz ($\lambda\simeq 1\,$cm), $W=100\,$MHz, $K=256$, $\Ptx=20\,$dBm, $\Gt=10\,$dB, $\Gr=2\,$dB, receiver noise figure $F=3\,$dB.

The metaprism is composed of $N \times M$ elements with spacing $d_x=d_y=\lambda/2$. 
For what the wall is regarded, it is supposed to be composed of areate concrete with \ac{e.m.} parameters $\epsilon_r=2.26$,  $\tan \delta=0.0491$ \cite{CorFra:94}, and scattering parameters $S^2=0.1$  (scattering coefficient), $R^2=0.9$  (reduction factor) \cite{PasMolMarRodSau:16}.
 
All locations on the left-side of the scenario are in \ac{NLOS} conditions because the signal from the \ac{BS} is obstructed by the upper-side building. As a consequence, the only possibility users located in that area have to communicate with the \ac{BS} is through specular/diffuse  scattering from the wall and/or through the metaprism, when present. 

In Fig. \ref{Fig:SNRNoMeta}, the \ac{SNR} map obtained computing \eqref{eq:SNR} in a dense grid of locations in the absence of metaprism.
The presence of the specular component caused by the wall oriented towards direction $\Theta=(-45^{\circ},0)$, i.e., obeying the Snell's law,  is evident. Users located along this direction can benefit to establish a communication with the \ac{BS}. In addition, the diffuse scattering component can in principle be exploited for communication as well, even though the corresponding \ac{SNR} values are in general low.

The impact of the metaprism on the \ac{SNR} map can be appreciated in  Fig. \ref{Fig:SNRMeta}. The phase profile of the metaprism has been designed according to the criterium illustrated in Sec. \ref{Sec:DesignBeamsteering} (beamsteering) by setting $\theta_m=40^{\circ}$ (note that $\thetai=45^{\circ}$). The map is shown for some values of subcarrier index $k$ corresponding to different colors, as indicated. The frequency-selective behavior of the metaprism allows to cover a wide range of angles in $\left [\theta_0^{(1)}= -25^{\circ}, \theta_0^{(K)}=-85^{\circ} \right ]$, corresponding to the ``prism effect". This effect can be better  exploited by the \ac{BS} by assigning to the generic user located at angle $\theta_0$ the subcarrier at which the metaprism reflects the signal towards $\theta_0$. The corresponding \ac{SNR} values are in general much higher than that obtained by exploiting only the reflections from the wall (see Fig. \ref{Fig:SNRNoMeta}).   
 
In Fig. \ref{Fig:SNRFocusing}, the \ac{SNR} map is shown when the criterium in Sec.  \ref{Sec:DesignFocusing}    (focusing) is used to design the phase profile of the metaprism by fixing the minimum focal distance $d_m=2\,$meters, and $\theta_m=5^{\circ}$. The map is shown for some values of subcarrier index $k$ corresponding to different colors. Differently from Fig. \ref{Fig:SNRMeta}, here it is evident that for each subcarrier the signal is more concentrated at a spefic location (focal distance), especially close to the metaprism. 
When moving from $k=K$ down to $k=1$, the focusing distance increases from $d_m$ to very large values, thus degenerating in beamsteering with angle $\theta_0=-\thetai \simeq -45^{\circ}$. 

We now investigate the impact of metaprism in terms of achievable rate \eqref{eq:Ru} supposing $U$ users are randomly located in the \ac{NLOS} square area $x\in [-15,-5]\,$meters, $z\in [2,10]\,$meters. The subcarrier assignment algorithm described in Sec. \ref{Sec:AR} is used. The plots in Fig. \ref{Fig:mAR} refer to the per-user achievable rate when increasing the number $U$ of total users in the area when different metaprism sizes $L_x=L_y=L$  are considered. As expected, when increasing the size of the metaprism the achievable rate increases thanks to the more favorable path-loss. 

In the absence of metaprism (wall), only the small percentage of users  fall within the small area corresponding to the specular reflection from the wall (at angle $\theta_0=-45^{\circ}$) or where the diffuse component is significant (see Fig. \ref{Fig:SNRNoMeta}). Most of users experience a bad \ac{SNR} condition.
Therefore, when increasing the number of users, the total available transmitted power is no longer sufficient to guarantee the same achievable rate to all users with a significant level. A way out is to change the allocation policy by satisfying only the few users in good \ac{SNR} condition and discarding all the others. In any case, the coverage of the system is in general very low. 
On the contrary, when introducing the metaprism, the performance improvement is very significant, at least a factor 5 with respect to the absence of the metaprism (only wall), even using metaprisms with reasonable dimension (e.g., $L=50\,$cm). The decreasing behavior of the plots is due to the fact that when increasing the number of users the total transmitted power is shared among more users included those experiencing bad \ac{SNR} conditions which require more power (higher weight $\omega^{(k)}$) to counteract their \ac{SNR} penalty (the policy in Sec. \ref{Sec:AR} imposes all users have the same achievable rate).
Beyond a certain value of $U$, depending on the size of the metaprism,  the per-user achievable rate drops to zero because the users in bad \ac{SNR} condition drain all the available power and hence it is no longer possible to guarantee the same achievable rate at a reasonable level.
As before, by discarding these disadvantaged users, one can maintain high values of achievable rate for the other users. 
In any case, the coverage obtained using the metaprism is significantly increased, as it can be deduced from the \ac{SNR} map in  Fig. \ref{Fig:SNRMeta}. 
 
When moving from a metaprism with dimension $L=50\,$cm to a metaprism with dimension $L=100\,$cm, an interesting phenomenon can be observed. First, there is a slightly decreasing of the achievable rate and the curve dropping happens at lower values of $U$.  This can be ascribed to the fact that by increasing the dimension of the metaprism, the corresponding power radiation patterns become more and more angle-selective. Then, the main lobes of the radiation patterns related to  two adjacent subcarriers tend to be less overlapped thus creating an angle gap between them which is not covered by the metaprism. All users with angles falling in these gaps experience low \ac{SNR} values. This problem can be overcome by increasing the number $K$ of subcarriers or decreasing the total bandwidth $W$. Again, different subcarrier assignment policies would bring to different behaviors.

Finally, we analyze the performance of focusing strategies when applied to a mono-dimensional scenario. In particular, we suppose the \ac{BS} is located at $20\,$meters from the metaprism with angle $\thetai=0^{\circ}$, where $U$ users are deployed randomly along the boresight direction of the metaprism in the range $[2,10]\,$meters. The beamsteering and focusing phase profiles in Sec. \ref{Sec:DesignBeamsteering} and  \ref{Sec:DesignFocusing}, respectively, are compared in terms of  achievable rate using the same subcarrier assignment algorithm.
The following parameters have been used during the design $d_m=2\,$meters, $\theta_m=0^{\circ}$. 

From the plots in Fig. \ref{Fig:ARMono}, it can be observed that by designing the metaprism to perform focusing there is a valuable performance improvement of 30-50\% with respect to a design based on beamsteering. The motivation is that within  this distance range the users are located in the Fresnel region (near-field) (see Fig. \ref{Fig:FDistance}), and hence they experience a path-loss advantage because of better signal concentration on user position (see Fig. \ref{Fig:PL2}) if the metaprism is designed to realize focusing.

\section{Conclusion}
\label{Sec:Conclusion}

In this paper, we have put forth the idea of \emph{metaprism}, a  metasurface designed with a frequency-dependent phase profile such that its reflection properties are dependent on the subcarrier index when illuminated by an \ac{OFDM}-like signal. 

Unlike \acp{RIS}, metaprisms are full passive (no energy supply needed) and they do not require a dedicated control channel to change their reflection properties, thus making them very appealing in all those situations  where low cost, easy of deployment, and back compatibility with existing radio interfaces are required when extending the  coverage  of \ac{NLOS} areas. Furthermore, they can work without the need for estimating the \ac{CSI} of both \ac{BS}-metasurface and metasurface-receiver links, which is one of the main current challenges of \acp{RIS}. 

We have provided design criteria for the phase profile of the metasprism to obtain subcarrier-dependent beamsteering and focusing functionalities within the area of interest.  In addition, we have proposed an example of low-complexity subcarrier assignment algorithm capable of guaranteeing all users  the same achievable rate.

The numerical results have put in evidence the significant improvement, in terms of coverage and achievable rate, which can be obtained using metaprisms, with respect to the situation where radio coverage is delegated to the natural specular and diffuse reflection from walls. In particular, the examples provided show an achievable rate increase of a factor of 5 and more, even using relative small-size metaprisms ($50\,$cm). 
We have also pointed out that, when operating at high frequencies (e.g., millimeter waves and beyond),  the near-field region of the \ac{e.m.} field is likely to extend to dozens meters  from the metaprism, making classical beamsteering-based design less efficient than focusing-based design.    

Future works will be devoted to the investigation of more complex networks including several \acp{BS} to evaluate the advantages of metaprisms  in terms of interference reduction. 
Another area of research could be related to the design of metasurfaces technologies tailored to the frequency-dependent characteristics introduced in this paper.


\ifCLASSOPTIONcaptionsoff
\fi
\bibliographystyle{IEEEtran}
\bibliography{IEEEabrv,MassiveMIMO,MetaSurfaces,PassiveRelays,WINS-Books}
\end{document}